\documentclass
[onecolumn]{revtex4-1}
\usepackage{subfigure}
\usepackage{amssymb,amsthm,amscd, amsbsy, array}
 \usepackage{amsmath,bm,color}
\usepackage{graphicx} 
\newcommand{\mathsym}[1]{{}}

\usepackage{epsfig}
\usepackage{epsf}    
\usepackage{rotating}
\usepackage{color}

\newcommand{\lf}{\left (}

\newcommand{\rg}{\right )}

\def\smallover#1/#2{\hbox{$\textstyle{#1\over#2}$}}
\def\beq{\begin{equation}}
\def\eeq{\end{equation}}
\def\beq{\begin{equation}}
\def\eeq{\end{equation}}
\def\bea{\begin{eqnarray}}
\def\eea{\end{eqnarray}}

\def\lf{\left(}
\def\rg{\right)}

\def\p{{\partial}}
\def\p{{\partial}}

\def\*{{\star}}

\def\aa{{{\`a }}}

\usepackage[authormarkup=none, markup=default, deletedmarkup=xout]{changes}

\definechangesauthor[color=purple]{3}
\definechangesauthor[color=blue]{vito}
\definechangesauthor[color=red]{1}
\begin{document}

 \title{Skyrmion tubes in achiral nematic liquid crystals}
\author{G. De Matteis \dag \ddag\S \footnote{e-mail: giovanni.dematteis@unisalento.it},  $\quad$  L. Martina \dag \ddag \footnote{e-mail: martina@le.infn.it}, $\quad$ C.  Naya \P \ddag\footnote{e-mail: carlos.naya@fysik.su.se}, $\quad$ V. Turco \dag \ddag\S \footnote{ e-mail: vito.turco@le.infn.it} 
\\ \dag Dipartimento di Matematica e Fisica, Universit\aa del Salento, Via per Arnesano, C.P. 193 I-73100 Lecce, Italy\\  \ddag  INFN, Sezione di  Lecce, Via per Arnesano, C. P. 193 I-73100 Lecce, Italy\\ \S GNFM-INDAM, Citt\`a Universitaria - P.le Aldo Moro 5, C. P. 00185 Roma, Italy \\
\P Department of Physics, Stockholm University, AlbaNova University Center, 106 91 Stockholm, Sweden\\}

\date{\today}

\begin{abstract}
We analyze the interaction with uniform external fields of nematic liquid crystals within a recent generalized free--energy posited by Virga and falling in the class of
quartic functionals in the spatial gradients of the nematic director.
We review some known interesting solutions, {\emph{i. e.}}, uniform heliconical structures, which correspond to the so-called twist-bend nematic phase and we
also study the transition between this phase and the standard uniform nematic one.
The twist-bend phase is further reproduced by 3D simulations. Moreover, we find liquid crystal configurations,
which closely resemble some novel, experimentally detected,  structures called Skyrmion tubes. Skyrmion tubes are characterized by a localized cylindrically-symmetric pattern
surrounded by either twist-bend or uniform nematic phase.
We study the equilibrium differential equations
and find numerical solutions and analytical approximations.
\end{abstract}
  
\maketitle
 
\section{Introduction}

The achiral nematic $N$ phase is surely the most common state for thermotropic liquid crystals. Due to the uniaxial symmetry of the constituent molecules, it is possible to describe this liquid crystalline phase by means of a director field $\bm{n}\in \mathbb{R}\text{P}^2$, prescribing point by point the average orientation of the molecular axes. In particular, in the $N$ phase the ground state alignment of these axes is parallel to a fixed direction $\bm{n}\equiv\bm{n}_0$. On the other hand, in the chiral nematic $N^*$ phase, formed by enantiomorphic molecules, the minimum energy configuration $\bm{n}=\bm{n}_c(\bm{r})$ is spontaneously twisted in a right-angle helix, whose pitch $P$ usually lies in the range of  micrometers. 

However, new classes of nematics are expected when molecules are less symmetric, as for example, the biaxial nematic phase \cite{Book2015, biaxial_universal, landaubiax}. Moreover, for strongly bent mesogenic molecules, a new modulated nematic phase, now recognized as the twist-bend nematic $N_{\text{TB}}$ phase, has been recently observed and reported in several works, starting from the breakthroughs in \cite{Chen2013,Borshch2013, vij2013}. It turned out that this phase is stabilized below the usual $N$ phase and, although formed by achiral molecules, it exhibits doubly degenerate chirality, consisting of right and left Meyer's heliconical domains \cite{Meyer76}. Thus, the appearance of the $N_{\text{TB}}$ phase represents a particularly interesting case of spontaneous breaking of the chiral symmetry. Since in these structures the director $\bm{n}$ is tilted by a fixed angle $0<\theta_0<\pi/2$, they may look similar to the smectic $SmC^*$ phases. However, at variance with them, the heliconical textures do not possess any layer periodicity. Moreover, the helical pitch is much smaller than the cholesteric one, \emph{i. e.}, of the order of $10\; \text{nm}$ \cite{Chen2013,Borshch2013}.   

Several papers addressed the theoretical analysis of the $N_{\text{TB}}$ phase, both from the phenomenological and the static continuum theory points of view \cite{selinger2013,Greco,Longa,Virga2}. More specifically, in \cite{Greco} a $N$-$N_{\text{TB}}$ phase transition was described by means of a generalized Maier-Saupe molecular field theory. In \cite{Longa}, a generalized Landau - de Gennes theory was applied to investigate the modulated nematic phases, possibly generated by  achiral and intrinsically  chiral bent mesogenic molecules. In \cite{Virga2}, the $N_{\text{TB}}$ phase was studied as a mixture of two different ordinary $N$ phases, both presenting heliconical structures with opposite helicities. A quadratic elastic theory, still featuring four Frank elastic moduli, was used for both helical phases. Similar models were proposed in \mbox{\cite{Barbero1,Barbero2, Barbero3}}, where also the effects of an external magnetic or electric bulk field were investigated.  Moreover, authors in \cite{VirgaTandF,Dozov2016} presented coarse-grained elastic models which, similarly to the model for $SmA^*$ \cite{DeGennes}, make use of an extra scalar order parameter. 

The prediction of the $N_{\text{TB}}$ phase dates back to the seminal paper \cite{Dozov2001} by Dozov, in which an elastic instability model was proposed with a bend constant $K_{33}$ turning negative. Higher derivative terms were added to the standard Frank-Oseen elastic energy in order to bound the energy from below. According to Dozov's model, depending on the ratio of $K_{11}$ and $K_{22}$ in the high-temperature non-modulated nematic regime, the low-temperature nematic phase can show either the twist-bend modulation when $K_{11}>2K_{22}$ or the splay-bend modulation otherwise. The same model predicted the existence of a nematic phase with spontaneous bend distortions \cite{Dozov2001,Meyer76}. However, unlike a twist, a pure bend distortion cannot fill the space without introducing frustration, possibly relieved by defects. This is certainly not the case for the mixed twist-bend distortion. Indeed, in \cite{Virga4} it was shown that in three space dimensions there exist only two families of director configurations which have uniform non-zero distortion characteristics at any point in space. It turned out that these latter configurations correspond to the right and left Meyer's heliconical domains, which form the $N_{\text{TB}}$ phase. Any other director field, apart from the constant nematic director, would be geometrically frustrated and become nonuniform if requested to fill the whole space. 

The natural successive step was to see whether it is possible to build an elastic free--energy that penalizes the departures from one of these uniform director fields. Since in the uniform heliconical phases only one of the distortion characteristics vanishes, namely the splay one, Frank's quadratic theory is no longer sufficient. Thus, a higher-order elastic theory, in which the bend elastic constant may turn negative, was proposed in \cite{Virga4}, allowing for fourth-order powers of $\nabla\bm{n}$ in the free--energy. The author focused on an achiral scenario where the $N_{\text{TB}}$ phase has been experimentally identified and deliberately built his generalized elastic free--energy with the symmetry of the intended  heliconical ground state, \emph{i. e.}, its double degeneration for right and left helicities. This choice makes the free--energy depend on only six elastic constants: three for the quadratic part and three for the quartic one. Then, for suitable choices of the elastic constants, it was shown how either the standard nematic or the heliconical phases minimize the proposed higher order free--energy. More specifically, the theory predicts the $N_{\text{TB}}$ phase arising from the standard nematic one for sufficiently negative values of the bend constant, passing through an intermediate pure bend state. 
 
In \cite{pre102us} we reviewed the theory presented in \cite{Virga4} and found that, in the same region where $N_{\text{TB}}$ is preferred, localized excitations of the heliconical ground state are possible. In particular, we showed how axisymmetric structures, with a radial dependence of the conical angle and an additional twist around the heliconical axis, are admissible states of the generalized elastic theory, with energies falling in between those of the heliconical ground state and the nematic alignment. We found that our soliton configurations resemble interesting axisymmetric structures recently observed in chiral nematics and chiral ferromagnets \cite{Rybakov2015,Du2018,PRB100,PRB98}, namely Skyrmion tubes. In contrast with these latter configurations, ours can be generated in an achiral framework without the need of external frustration.

Emergent topological defects in condensed matter systems are drawing much attention, particularly due to its potential technological applications, and Skyrmion tubes are not an exception. For instance, they have been recently proposed as magnonic waveguides channeling spin waves, based on the propagation of their breathing and rotational modes \cite{Xing2020}.
However, theoretical studies such as \cite{PRB100,PRB98} have been focused on its realization in ferromagnets, although its experimental attainment seems to be easier in liquid crystals \cite{PRB100}. In this context, \cite{pre102us} paved the way for a better theoretical understanding of Skyrmion tubes in liquid crystals, where other localized configurations such as helicoids or Skyrmions are well known
\cite{Fukuda2011,Ackerman2014,Leonov2014,Afghah2017,DeMatteis2018,DeMatteis2019, DeMatteis2020JOI, DeMatteis2019_MCLC, DeMatteis2018PRE}. 

In the present paper, we generalize our previous work \cite{pre102us} with the addition of an external uniform magnetic field. We find Skyrmion-tube-like nematic textures that form when a uniform magnetic field
is applied along the axis of a heliconical state. These Skyrmion tubes are surrounded by either a nematic uniform phase or by a uniform twist-bend phase.
The paper is organized as follows. In Sec. \ref{secttwo}, we briefly revise previous material and set the model. Then, we study the interaction of an external magnetic field and find a uniform heliconical state. We also investigate the transition
to the standard uniform nematic phase as the magnitude of the external field is increased from zero to a critical value. In Sec. \ref{sectfour}, we study nonuniform localized states under the action of the external field by imposing
boundary conditions at the center of the heliconical state and at infinity. We find Skyrmion-tube configurations where the nematic texture is nonuniform in a localized radial region immersed in a uniform, either nematic or heliconical, state.
Finally, in Sec. \ref{sectfive} we draw our conclusions and outline future investigations.

\section{Twist-bend phase under external fields}
\label{secttwo}

Nematic liquid crystals are usually modeled by Frank's elastic free--energy density.
This is a general positive-definite quadratic form in the spatial gradients $\nabla\bm{n}$ of a unit vector, the nematic director $\bm{n}$,
and it is written as
\begin{equation}
\label{frank_energy_std}
F_{\text{F}} = \frac{1}{2}K_{11}(\text{div}\bm{n})^2+\frac{1}{2}K_{22}(\bm{n}\cdot\text{curl}\bm{n})^2+\frac{1}{2}K_{33}|\bm{n}\times \text{curl}\bm{n}|^2+
K_{24}\left[\text{tr}(\nabla\bm{n})^2-(\text{div}\bm{n})^2\right],
\end{equation}
where $K_{11}, K_{22}, K_{33}$, and $K_{24}$ are the Frank elastic constants and they are such that
\begin{equation}
K_{11}-K_{24}>0,\qquad K_{22}-K_{24}>0,\qquad K_{33}>0,\qquad K_{24}>0,
\end{equation}
known as Ericksen's inequalities \cite{ericksen1966}.
The term $K_{24}$ is a {\emph{null Lagrangian}},
it can be integrated over the domain $\mathcal{B}$ occupied by the nematic medium without producing any contribution
to the total free--energy, provided that $\bm{n}$ is assigned over the boundary $\partial\mathcal{B}$.

In \cite{Selinger2018} it was shown that
Frank's elastic free--energy density can be written as a quadratic form in four quantities $(S,T,\bm{b},\mathbf{D})$ as follows
\begin{eqnarray}
\label{frank_free_energy}
F_{\text{F}} = \frac{1}{2}\left(K_{11}-K_{24}\right)S^2 + \frac{1}{2}\left(K_{22}-K_{24}\right)T^2+\frac{1}{2}K_{33}B^2+K_{24}\text{tr}(\mathbf{D}^2),
\end{eqnarray}
where $S=\text{div}\bm{n}$ is a scalar called {\emph{splay}}, $T=\bm{n}\cdot\text{curl}\bm{n}$ is a pseudo-scalar named {\emph{twist}}, and $B^2=\bm{b}\cdot\bm{b}$, with the vector
$\bm{b}=\bm{n}\times\text{curl}\bm{n}$ being the so-called {\emph{bend}}. $\mathbf{D}$ is a symmetric traceless tensor such that $\mathbf{D}\bm{n}=\bm{0}$.
Accordingly, it can be given the form
\begin{equation}
\mathbf{D}=q(\bm{n}_1\otimes\bm{n}_1-\bm{n}_2\otimes\bm{n}_2),
\end{equation}
where $q$ is the positive eigenvalue of $\mathbf{D}$, named by Selinger \cite{Selinger2018} as {\emph{biaxial splay}},
and $\bm{n}_1$ and $\bm{n}_2$ are the eigenvectors orthogonal to $\bm{n}$. 
The tensor $\mathbf{D}$ can also be given the following form in terms of $\nabla\bm{n}$
\begin{eqnarray}
D_{ij} = \frac{1}{2}\left[\partial_in_j+\partial_jn_i-n_in_k\partial_kn_j-n_jn_k\partial_kn_i-\delta_{ij}\text{div}\bm{n}+n_in_j\text{div}\bm{n}\right].
\end{eqnarray}
The quantities $(S,T,\bm{b},\mathbf{D})$ are independent from one another and are called {\emph{measures of distortion}}. Frank's energy (\ref{frank_free_energy}) admits as global minimizer the state
\begin{equation}
\label{UniformState}
S = T = B = q = 0,
\end{equation}
which corresponds to any constant field $\boldsymbol{n} \equiv \boldsymbol{n_0}$.

In \cite{Virga4} it was put forward a new energy functional with quartic powers of measures of distortion $(S, T, \bm{b},\mathbf{D})$ as follows
\begin{equation}
\label{virga_free_energy}
F_{TB}(S,T, \bm{b},\mathbf{D})=\frac{1}{2}k_1S^2+\frac{1}{2}k_2T^2+k_2\text{tr}\mathbf{D}^2+\frac{1}{2}k_3B^2+\frac{1}{4}k_4T^4+k_4(\text{tr}\mathbf{D}^2)^2+\frac{1}{4}k_5B^4+
k_6T\bm{b}\cdot\mathbf{D}(\bm{n}\times\bm{b}).
\end{equation}
This represents the lowest order free--energy density that, for a suitable choice of the elastic constants, admits as global minimizer the so-called heliconical uniform distortion state \cite{Virga4, pre102us}, 
as opposed to the uniform state (\ref{UniformState}).
By directly comparing (\ref{virga_free_energy}) with (\ref{frank_free_energy}) we get the following formal identification
\begin{equation}
\label{el_consts_identif}
k_1=K_{11}-K_{24},\quad k_2=K_{22}-K_{24}=K_{24},\quad k_3=K_{33},
\end{equation}
but as shown below $k_3$ can also assume negative values.
From (\ref{el_consts_identif}), it is clear that the number of independent Frank elastic constants is reduced from four to three as $K_{22}-K_{24}=K_{24}$.
This assumption is due to the choice of the same elastic constant in front of $T^2$ and $2\text{tr}\mathbf{D}^2$ in the quadratic part of (\ref{virga_free_energy}), and $T^4$ and $4(\text{tr}\mathbf{D}^2)^2$
in the quartic part.
This latter condition is related to the heliconical global minimizer of (\ref{virga_free_energy}) (see below), which is such that $T^2=2q^2$ \cite{Virga4, pre102us}. Hence,
the free-energy density must be invariant under the transformation $T^2\leftrightarrow 2q^2$, implying that only the combinations $T^2+2q^2$
and $T^4+4q^4$ appear in the free-energy density. 

\begin{figure}[ht]
\begin{center}
\includegraphics[width=0.45\textwidth]{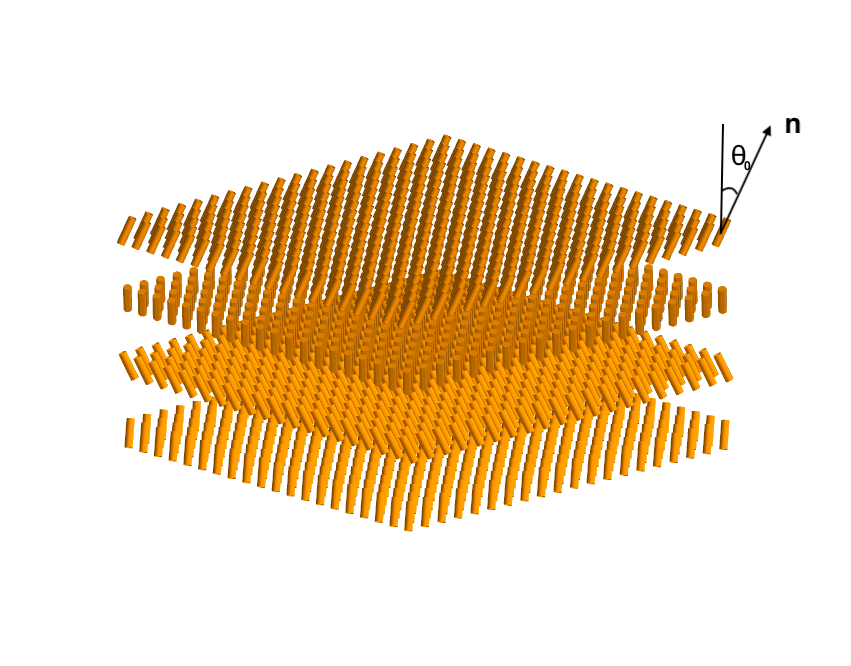}
\includegraphics[width=0.45\textwidth]{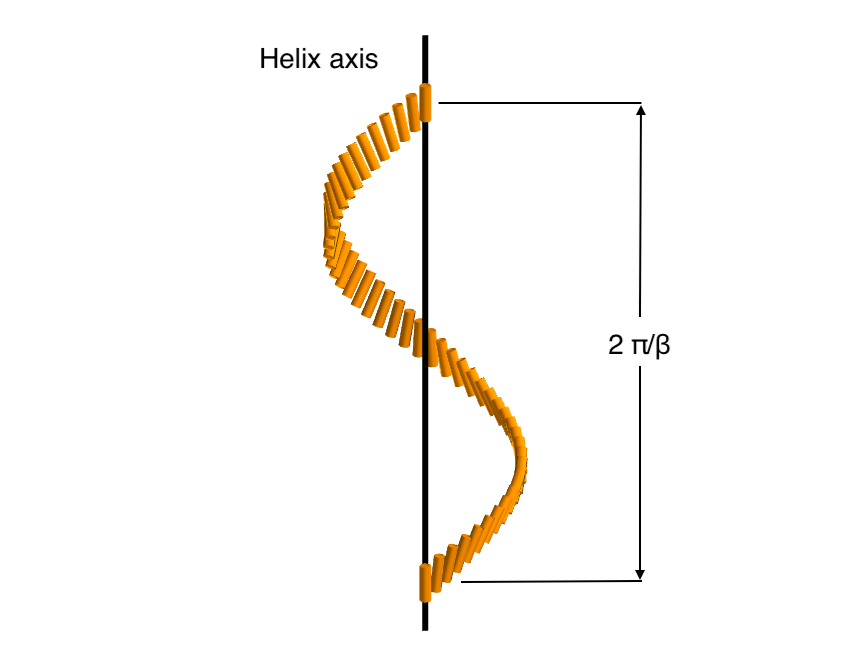}
\end{center}
\caption{Three-dimensional representation of the uniform heliconical distortion. {\it Left:} Different $(x,y)$-plane cross sections showing the change of orientation along the $z$ direction. {\it Right}: Helix line along the $z$ axis for a fixed
distance from it.}
\label{3D-alhpa0}
\end{figure}

The above energy density turns out to be coercive provided that
\begin{equation}
\label{constraintsvirga2paperh}
k_4>0,\quad k_5>0,\quad k_6>0,\quad k_6^2<2k_4k_5,
\end{equation}
which is the condition of positive definiteness of the quartic part of (\ref{virga_free_energy}).
In terms of $\bm{n}$ and its gradients $\nabla\bm{n}$, (\ref{virga_free_energy}) can be written as follows \cite{pre102us}
\begin{eqnarray}
\label{virga_free_energy2}
F_{TB}&=&\frac{1}{2}(k_1-k_2)(\text{div}\bm{n})^2+k_2(\bm{n}\cdot\text{curl}\bm{n})^2+k_2\text{tr}(\nabla\bm{n})^2+\frac{1}{2}k_3|\bm{n}\times\text{curl}\bm{n}|^2+
\frac{1}{4}k_4(\bm{n}\cdot\text{curl}\bm{n})^4\nonumber\\
&+&
k_4\left[\text{tr}(\nabla\bm{n})^2
+\frac{1}{2}(\bm{n}\cdot\text{curl}\bm{n})^2-\frac{1}{2}(\text{div}\bm{n})^2\right]^2+\frac{1}{4}k_5|\bm{n}\times\text{curl}\bm{n}|^4
\nonumber\\
&-&k_6\left[(\bm{n}\cdot\text{curl}\bm{n})\text{curl}\bm{n}\cdot(\nabla\bm{n})(\bm{n}\times\text{curl}\bm{n})+
\frac{1}{2}(\bm{n}\cdot\text{curl}\bm{n})^2|\bm{n}\times\text{curl}\bm{n}|^2\right].
\end{eqnarray}
Correspondingly, the free--energy stored in a region $\mathcal{B}$ occupied by the liquid crystal is given by the volume integral
\begin{equation}
\label{free_energy_volume}
\mathcal{F}=\int_{\mathcal{B}}{F_{TB}\text{d}\mathcal{B}}.
\end{equation}
As mentioned above, (\ref{virga_free_energy2}) admits, as global minimizer, the uniform heliconical state \cite{Virga4, pre102us}. This latter can be written as follows
\begin{equation}
\label{heliconics_1}
\bm{n}_h=\sin\theta_0\cos\beta z\bm{e}_x+\sin\theta_0\sin\beta z\bm{e}_y+\cos\theta_0\bm{e}_z,
\end{equation}
where $\bm{e}_x,\bm{e}_y,\bm{e}_z$ are the Cartesian unit  basis vectors in $\mathbb{R}^3$, $\theta_0$ is the {\emph{conical}} angle and $\beta$ a parameter that provides the pitch $P=2\pi/|\beta|$ of the twist.
Here, we assume that $\beta$ is a characteristic parameter that depends on the elastic constans $k_i$ only and needs to be optimized.
The three-dimensional representation of such configurations is displayed in Fig. \ref{3D-alhpa0},
where a set of $(x,y)$-plane cross sections showing how the configuration changes along $z$ and a specific helix line are depicted.
The nematic director $\bm{n}_h$ rotates around $\bm{e}_z$ making a fixed cone angle $\theta_0$ with the rotation axis $\bm{e}_z$, which is called the helix axis.
The structure (\ref{heliconics_1}) describes therefore the heliconical distortion predicted by Meyer \cite{Meyer76} and it corresponds to the {\emph{twist-bend}}
liquid crystal phase $N_{\text{TB}}$, experimentally detected in 2011 \cite{cestari}. It is worth noticing that formula (\ref{heliconics_1}) also describes  the nematic phase $N$
when $\theta_0=0$ and the chiral nematics when $\theta_0=\frac{\pi}{2}$, implying that the twist-bend phase represents a structural link between these two extreme phases.
Of course, as also observed in \cite{Virga4}, the heliconical configurations cannot be minimizers of the standard Frank elastic energy,
and a new elastic theory as (\ref{virga_free_energy}) was needed to accommodate the heliconical phase as a ground state.

The corresponding free--energy density reads
\begin{eqnarray}
\label{freeenergydensity_nofield}
F_{TB}(\bm{n}_h)=f_{TB}(\theta_0,\beta)&=&\frac{1}{8}\left(-4k_6\cos^2\theta_0\sin^6\theta_0+4k_4\sin^8\theta_0+\frac{1}{8}k_5\sin^42\theta_0\right)\beta^4\nonumber\\
&+&\frac{1}{8}\left(8k_2\sin^4\theta_0+k_3\sin^22\theta_0\right)\beta^2,
\end{eqnarray}
which depends on the pitch-related parameter $\beta$ and the conical angle $\theta_0$, and is minimized by the values
\begin{equation}
\label{beta}
\beta=\pm\frac{(2k_2k_5+k_3k_6)+2(k_3k_4+k_2k_6)}{\sqrt{-(2k_2k_5+k_3k_6)(2k_4k_5-k_6^2)}},
\end{equation}
where $\pm$ signs label a counterclockwise or a clockwise heliconical configuration, and
\begin{equation}\label{def_theta_0}
\theta_0=\arcsin{\left(\sqrt{\frac{2k_2k_5+k_3k_6}{(2k_2k_5+k_3k_6)+2(k_3k_4+k_2k_6)}}\right)}.
\end{equation}
In \cite{Virga4, pre102us} it was shown that, in order to have the heliconical states \eqref{heliconics_1}, the following constraints on the elastic constants must hold
\begin{equation}
\label{elastic_constraints}
2 k_4 k_5-k_6^2>0,\qquad k_3 k_6+2k_5 k_2<0,\qquad k_2 k_6+k_3 k_4<0. 
\end{equation}
When an external magnetic field $\bm{H}$ is applied, the free--energy density (\ref{virga_free_energy2}) turns into
\begin{eqnarray}
\label{virga_free_energy2_mag}
F_{H}&=&\frac{1}{2}(k_1-k_2)(\text{div}\bm{n})^2+k_2(\bm{n}\cdot\text{curl}\bm{n})^2+k_2\text{tr}(\nabla\bm{n})^2+\frac{1}{2}k_3|\bm{n}\times\text{curl}\bm{n}|^2+
\frac{1}{4}k_4(\bm{n}\cdot\text{curl}\bm{n})^4\nonumber\\
&+&
k_4\left[\text{tr}(\nabla\bm{n})^2
+\frac{1}{2}(\bm{n}\cdot\text{curl}\bm{n})^2-\frac{1}{2}(\text{div}\bm{n})^2\right]^2+\frac{1}{4}k_5|\bm{n}\times\text{curl}\bm{n}|^4
\nonumber\\
&-&k_6\left[(\bm{n}\cdot\text{curl}\bm{n})\text{curl}\bm{n}\cdot(\nabla\bm{n})(\bm{n}\times\text{curl}\bm{n})+
\frac{1}{2}(\bm{n}\cdot\text{curl}\bm{n})^2|\bm{n}\times\text{curl}\bm{n}|^2\right]+\Gamma_{\rm H},
\end{eqnarray}
where we added the term
\begin{equation}
 \Gamma_{\rm H} = -\frac{\chi_a}{2}\left(\bm{n}\cdot\bm{H}\right)^2,
 \end{equation}
$\chi_a$ being the magnetic susceptibility of the liquid crystal material.
Otherwise, the free--energy can be also written as follows
\begin{eqnarray}
F_{H} &=& \frac{1}{2}k_1S^2+\frac{1}{2}k_2T^2+k_2\text{tr}\mathbf{D}^2+\frac{1}{2}k_3B^2+\frac{1}{4}k_4T^4+k_4(\text{tr}\mathbf{D}^2)^2+\frac{1}{4}k_5B^4
\nonumber \\
&+&k_6T\bm{b}\cdot\mathbf{D}(\bm{n}\times\bm{b})-\frac{\chi_a}{2}\left(\bm{n}\cdot\bm{H}\right)^2.
\end{eqnarray}
Correspondingly, the stored free--energy in a region $\mathcal{B}$ occupied by the liquid crystal is given by the volume integral
\begin{equation}
\label{free_energy_volume}
\mathcal{F}_{H}=\int_{\mathcal{B}}{F_{H}\text{d}\mathcal{B}}.
\end{equation}
The Euler-Lagrange equation associated with the above functional is given by
\begin{equation}
\label{euler_lagrange_eq_virga_f}
\frac{\partial F_H}{\partial \bm{n}}-\text{div}{\left(\frac{\partial F_H}{\partial\nabla\bm{n}}\right)}=\lambda\bm{n},
\end{equation}
where $\lambda$ is a Lagrange multiplier for the unit director constraint. To obtain a pure equation it suffices crossing by $\bm{n}$ both sides of (\ref{euler_lagrange_eq_virga_f}).

In the following, we will consider an external magnetic field along the $z$-axis, {\emph{i. e.}}, $\bm{H}=H\bm{e}_z$,
and assume a nematic director field as in (\ref{heliconics_1})
\begin{equation}
\label{heliconics_2_magnetic_field}
\bm{n}_h=\sin\theta_0\cos\beta z\bm{e}_x+\sin\theta_0\sin\beta z\bm{e}_y+\cos\theta_0\bm{e}_z,
\end{equation}
with the helix axis parallel to the magnetic field.
As mentioned above \cite{PRB100}, $\beta$ is to be taken as fixed by the elastic constants $k_i$ only in accordance with (\ref{beta}).
Thus, we assume that the external field just affects the nematic director by a torque
\begin{equation}
\label{magtorque}
\tau_H=-\chi_a (\bm{n}\cdot\bm{H})\bm{H}\times\bm{n},
\end{equation}
imparted to the liquid crystal molecules. Accordingly, the conical angle $\theta_0$ will change.
The interaction with an external field $\bm{H}=H\bm{e}_z$ is explicitely given in terms of $\theta_0$ by the term
\begin{equation}
 \Gamma_{\rm H} = -\frac{\chi_a}{2}\left(\bm{n}\cdot\bm{H}\right)^2=- \frac{1}{2} \chi_a H^2 \cos^2 \theta_0.
 \end{equation}
Correspondingly, the reduced free--energy density takes the form
\begin{equation}
\label{function_tH}
F_{H}(\bm{n}_h)=f_{H}(t)=\frac{1}{4}\left[-2k_6(1-t)t+2k_4t^2+k_5(1-t)^2\right]t^2\beta^4+\frac{1}{2}\left[2k_2t+k_3(1-t)\right]t\beta^2-\frac{1}{2}\chi_aH^2(1-t).
\end{equation}
where $t=\sin^2\theta_0$.
We need now to minimize (\ref{function_tH}) by solving the stationary condition
\begin{equation}
\partial_tf_{H}(t)=0,
\end{equation}
which is equivalent to (\ref{euler_lagrange_eq_virga_f}) under parametrization (\ref{heliconics_2_magnetic_field}).
The latter equation becomes
\begin{equation}
\label{eqtobestudied}
\beta^2\left(k_3+4k_2t-2k_3t\right)+\beta^4t\left[t\left(-3k_6+4k_4t+4k_6t\right)+k_5\left(1-3t+2t^2\right)\right]+H^2\chi_a=0,
\end{equation}
and the real solution to (\ref{eqtobestudied}) is given by 
\begin{equation}
t = \sqrt[3]{\sqrt{\Delta}-\frac{\eta}{2}} - \sqrt[3]{\sqrt{\Delta}+\frac{\eta}{2}} - \frac{b}{3a},
\end{equation}
where
\begin{equation}
\Delta = \left(\frac{\gamma}{3}\right)^3+\left(\frac{\eta}{2}\right)^2,\qquad \gamma=-\frac{1}{3}\left(\frac{b}{a}\right)^2+\frac{c}{a},\qquad \eta = \frac{2}{27}\left(\frac{b}{a}\right)^3-\frac{bc}{3a^2}+\frac{d}{a},
\end{equation}
with
\begin{equation}
a = \beta^4(2k_4+k_5+2k_6),\qquad b = -\frac{3}{2}\beta^4(k_5+k_6),\qquad c = \beta^2\left(2k_2-k_3+\frac{\beta^2}{2}k_5\right),\qquad d = \frac{\beta^2}{2}k_3+\frac{1}{2}H^2\chi_a.
\end{equation}
In order to have a unique real solution, the discriminant $\Delta$ must be positive. This is the case when the external field vanishes \cite{Virga4, pre102us}.
When $H\neq 0$, it is clear from the definition of the quantities $d$ and $\eta$ that $\Delta$ increases with respect to the zero-field value, thus keeping the positive sign
and still yielding a unique real solution.

\begin{figure}[ht]
\begin{center}
\includegraphics[width=.7\textwidth]{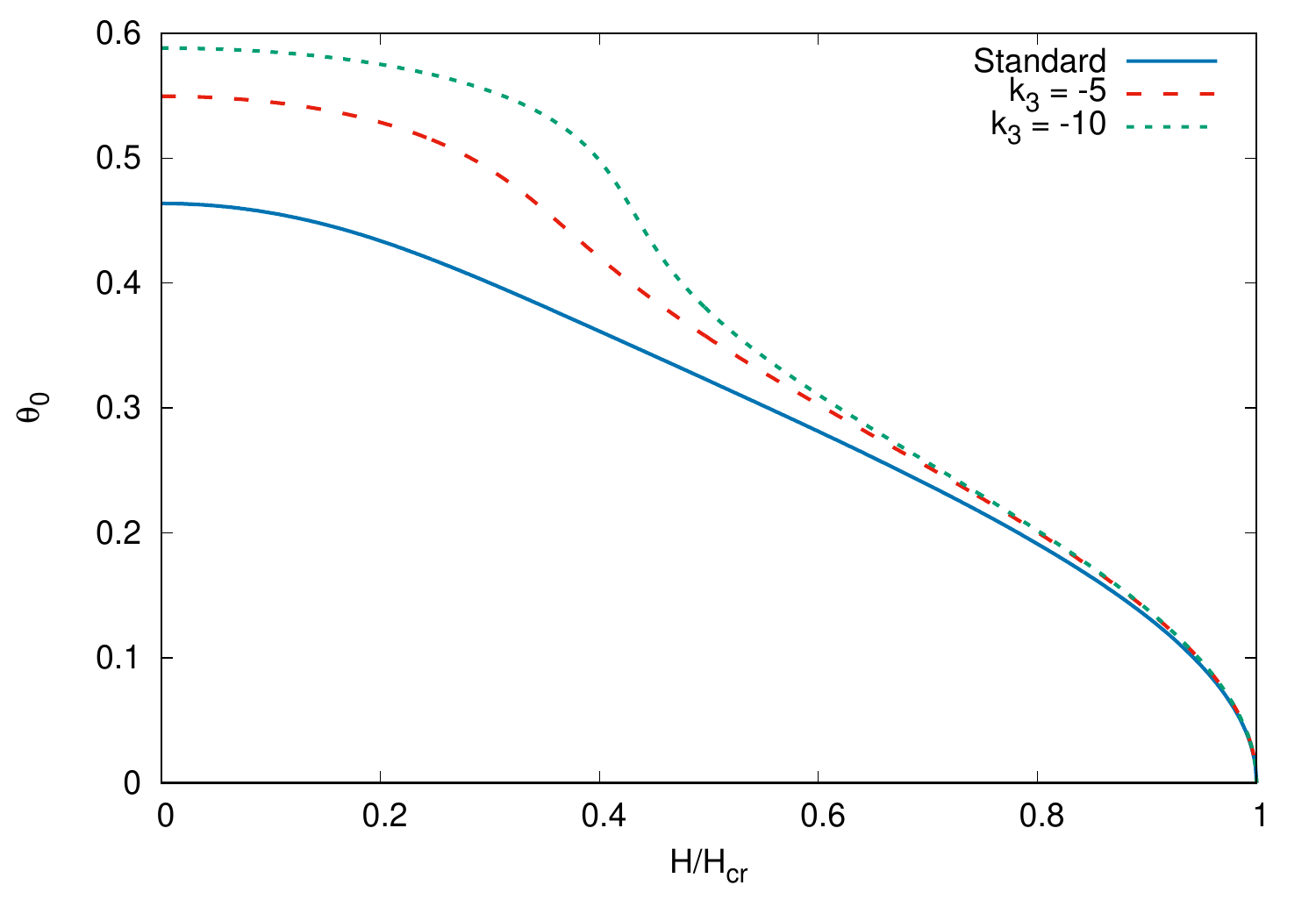}
\end{center}
\caption{Conical angle as a function of the external field upon changing elastic constant $k_3$: blue solid line for the standard set ($H_{\rm cr}^2 = 75/\chi_a$); red dashed line when $k_3 = -5$ ($H_{\rm cr}^2 = 605/3 \chi_a$); green dotted line for $k_3 = -10$ ($H_{\rm cr}^2 = 845/\chi_a$).}
\label{theta0-k3bis}
\end{figure}

\begin{figure}[ht]
\begin{center}
\includegraphics[width=.75\textwidth]{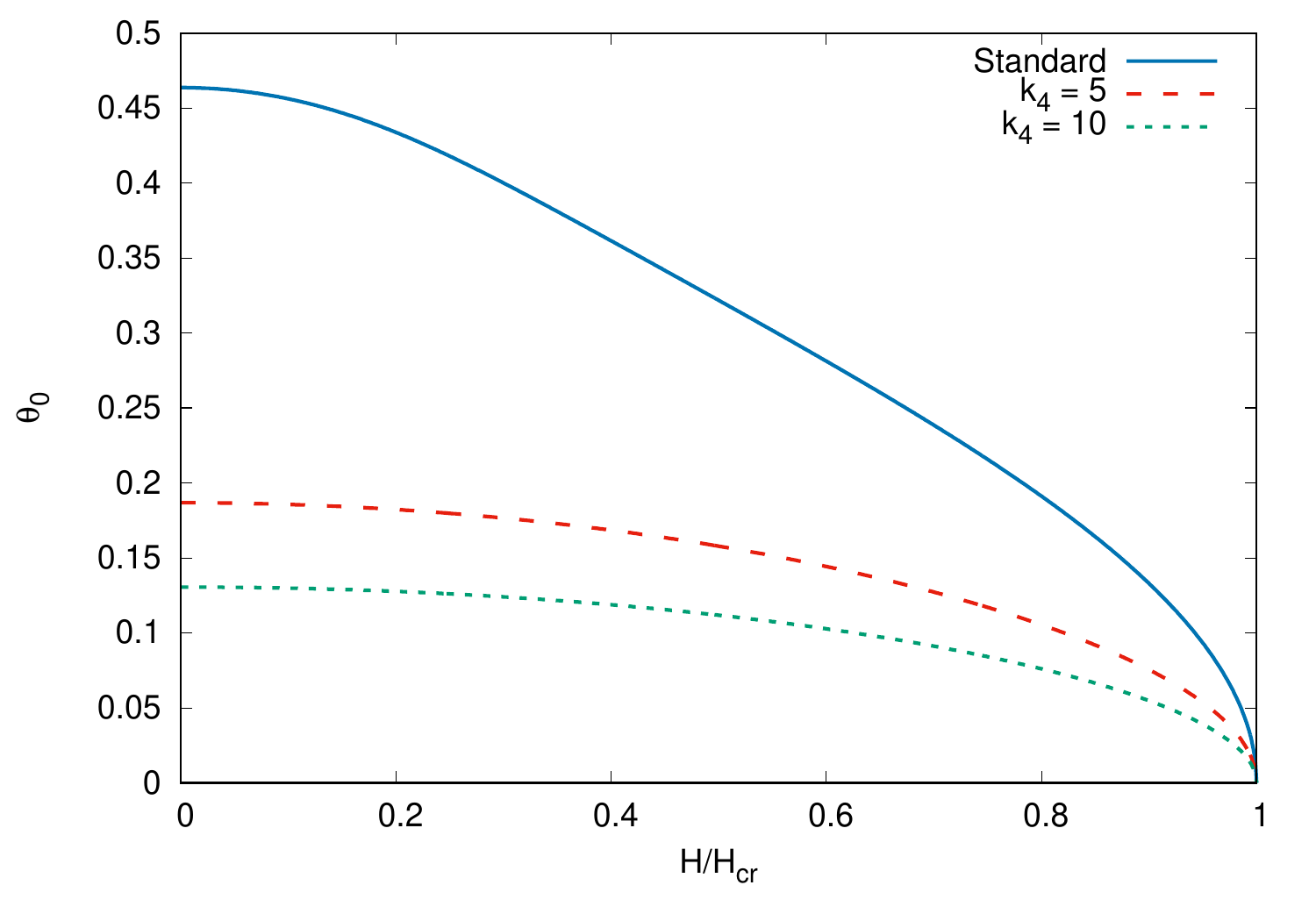}
\end{center}
\caption{Conical angle as a function of the external field upon changing elastic constant $k_4$: blue solid line for the standard set ($H_{\rm cr}^2 = 75/\chi_a$); red dashed line when $k_4 = 5$ ($H_{\rm cr}^2 = 841/3 \chi_a$); green dotted line for $k_4 = 10$ ($H_{\rm cr}^2 = 10443/19 \chi_a$).}
\label{theta0-k4bis}
\end{figure}

Therefore,
\begin{equation}
\label{thetazero_emf}
\theta_0=\arcsin{\sqrt{\sqrt[3]{\sqrt{\Delta}-\frac{\eta}{2}} - \sqrt[3]{\sqrt{\Delta}+\frac{\eta}{2}} - \frac{b}{3a}}},
\end{equation}
which generalizes (\ref{def_theta_0}) when an external field is present.
In addition, we also study the transition to the standard uniform nematic phase corresponding to $t=0$ where the director field lines up with the direction of the external field, \emph{i. e.}, along $\bm{e}_z$.
For this to occur, the external field should solve equation (\ref{eqtobestudied}) when $t=0$, {\emph{i. e.}},
\begin{equation}
\beta^2k_3+H^2\chi_a=0,
\end{equation}
which leads to the critical field
\begin{equation}
H_{\text{cr}} = \pm\sqrt{-\frac{\beta^2 k_3}{\chi_a}} .
\end{equation}
In the following, when dealing with an external field, we will give it in terms of this critical field, $H_{\rm cr}$.
Thus, when $H\geq H_{\text{cr}}$, $t=0$, that is to say, the phase is standard uniform nematic.
In terms of the elastic constants, the critical field becomes
\begin{equation}
\label{hcritical}
H_{\text{cr}} = \pm\sqrt{\frac{1}{\chi_a}\frac{k_3(2k_3k_4+2k_2k_5+2k_2k_6+k_3k_6)^2}{(2k_2k_5+k_3k_6)(2k_4k_5-k_6^2)}},
\end{equation}
which has been obtained by choosing for the parameter $\beta$ the expression in (\ref{beta}).
In Figs. \ref{theta0-k3bis} and \ref{theta0-k4bis} we represent the conical angle $\theta_0$ as a function of the external field for different choices of the elastic constants.
It is interesting to see in Fig. \ref{theta0-k3bis} how $\theta_0$ goes to zero when increasing $H$. In particular, we have plotted (blue line) the curves $\theta_0 (H)$ for 
the values of the elastic constants $k_1 = k_2 = k_4 = k_5 = k_6 = 1$ and $k_3 = -3$ (from now on we will call this choice of values the {\it standard set})
together with a decreasing of $k_3$ to -5 (red dashed line), which corresponds to a critical field $H_{\rm cr}^2 = 605/3 \chi_a$, and $k_3 = -10$ (green dotted line) with $H_{\rm cr}^2 = 845/\chi_a$. Interestingly, we see that in all cases the approach to 0, when we are close to the critical field, occurs in the same way. However, the behaviour for small values of the external field is quite different. When decreasing $k_3$, $\theta_0$ takes longer to become significantly smaller and a more abrupt reduction appears.
Similarly, we can consider the case of an increasing elastic constant $k_4$.
In Fig. \ref{theta0-k4bis}, besides the diminution of $\theta_0$ for the standard set of parameters, the cases of only changing $k_4$ from 1 to 5 and 10 (red dashed line and green dotted line, respectively) are shown. In this case, the critical field when $k_4 = 5$ is $H_{\rm cr}^2 = 841/3 \chi_a$ and $H_{\rm cr}^2 = 10443/19 \chi_a$ for $k_4 = 10$. Unlike the previous case, for an increasing $k_4$ the conical angle shrinks in a more regular way. This might be also favoured because, even in the absence of an external field, $\theta_0$ considerably decreases with $k_4$.

For small values of the ratio $h=\frac{H}{H_{\rm cr}}$, $t$ gets the form
\begin{equation}
t = t_0 + \frac{\beta ^2 k_3 \left(\sqrt[3]{2\sqrt{\Delta_0}-\eta _0}+\sqrt[3]{2\sqrt{\Delta_0}+\eta _0}\right)}{12\sqrt[3]{2} \;a\; \sqrt{\Delta_0}}  h^2 + O\left(h^4\right)
\end{equation}
where all quantities subindexed with $0$ refer to the corresponding quantities when $H=0$.
In terms of the asymptotic angle one has
\begin{equation}
\theta_0 = \arcsin{\left(\sqrt{\frac{2k_2k_5+k_3k_6}{(2k_2k_5+k_3k_6)+2(k_3k_4+k_2k_6)}}\right)} + \frac{1}{\sqrt{1-t_0}}
\frac{\beta ^2 k_3 \left(\sqrt[3]{2\sqrt{\Delta_0}-\eta _0}+\sqrt[3]{2\sqrt{\Delta_0}+\eta_0}\right)}{12\sqrt[3]{2} \;a\; \sqrt{\Delta_0}}\; h^2 +O(h^3).
\end{equation}
In this formula the dependency on $k_3$ is quite involved, but algebraic, then analytic.
Notice that in this model the response of the heliconical angle $\theta_0$ to the external field is just quadratic as in a sort of Kerr effect \cite{kittel}.

\begin{figure}[ht]
\begin{center}
\includegraphics[width=.75\textwidth]{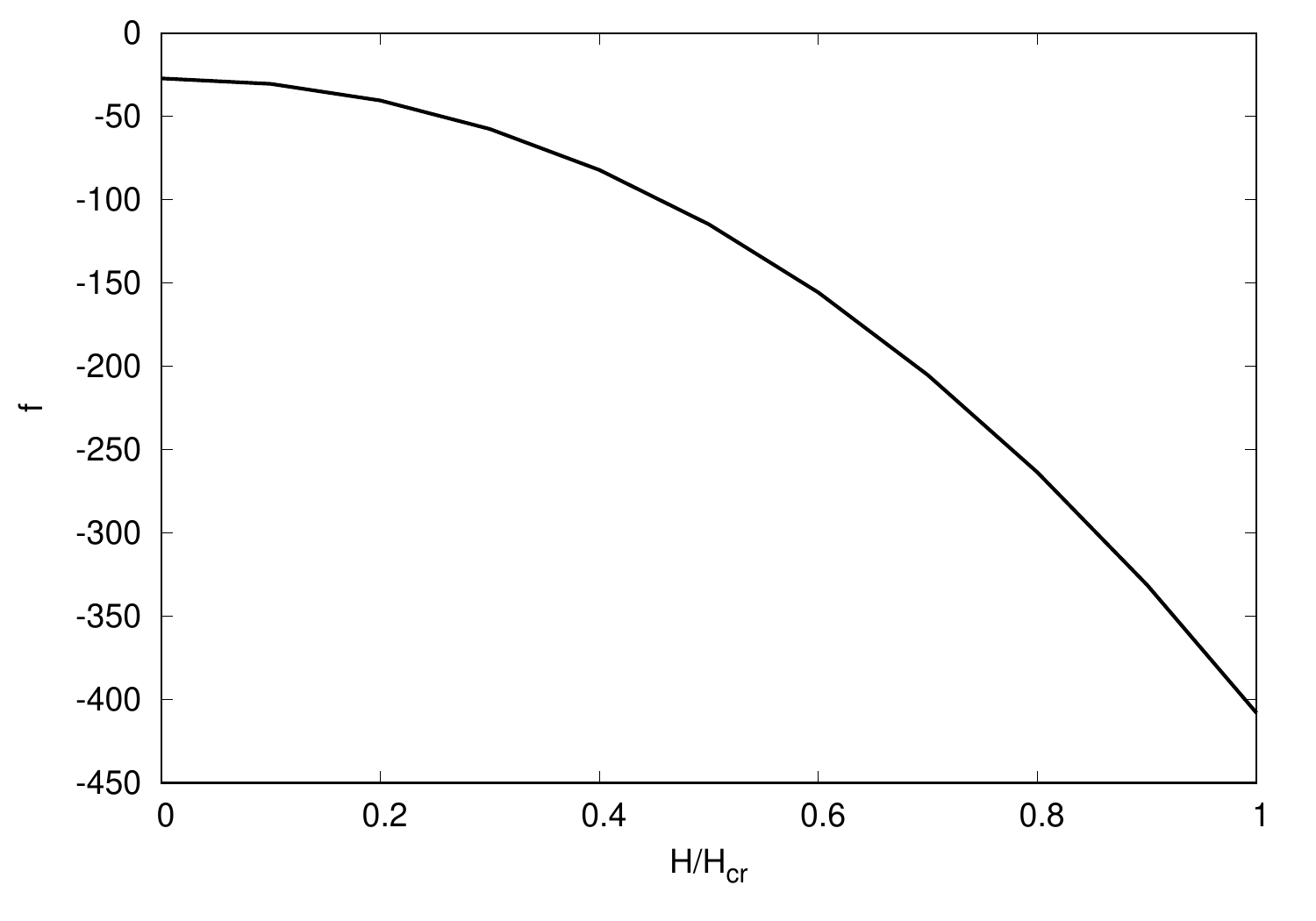}
\end{center}
\caption{Energy per volume of the twist-bend phase as a function of the external field for the standard set of parameters.}
\label{Energy-TB}
\end{figure}

Despite the complexity of the system under study, even when only considering the simple twist-bend configuration, simulations in 3 dimensions of this ground state have been successfully undertaken. One should note that within this theoretical setup, performing 3D simulations to minimize \eqref{free_energy_volume} via (\ref{euler_lagrange_eq_virga_f}) is a challenging problem. On the one hand, the boundary conditions for the vector director are not constant at infinity, together with a free energy which does not vanish asymptotically. In addition, a high accuracy in the numerical calculation is needed in order to exactly match the values obtained from the analytical study, not only for the conical angle $\theta_0$ but also for the pitch $\beta$. In the case of the twist-bend ground state, this may be done in a lattice of a reasonable size. In this way, it has been confirmed, for the standard set of parameters (with a lattice spacing of $0.02$ and a gradient flow method), the dependence of $\theta_0$ on the external field as in Fig. \ref{theta0-k3bis} (solid line), besides validating the assumption of constant pitch made before. Furthermore, in Fig. \ref{Energy-TB} one can see the energy per volume, $f$, as a function of $H/H_{\rm cr}$, where for $H = H_{\rm cr}$ we arrive at the nematic phase.

As mentioned in the introduction, several studies found that the pitch $P$ of the modulated nematic structure falls in the scale of a few nanometers \cite{Chen2013, Borshch2013}. The same studies report, under Freeze-Fracture Transmission Electron Microscopy (FFTEM), the presence of stripe-textured fracture planes which indicates fluid layers periodically arrayed in the bulk with a spacing of $P$. On the other hand, the authors in \cite{Chen2013} found that this periodic structure is achieved with no detectable associated modulation of the electron density, and so it is not accompanied
by a mass density wave, revealing a nematic rather than smectic molecular ordering. Thus, the “layers” in 3D found in the FFTEM are not images of molecular scale interfaces, but rather are 2D surfaces of constant azimuthal phase of the heliconical precession, sometimes called "pseudolayers" \cite{challa}. Due to the undulation of these surfaces, the direction of the heliconical axes periodically changes accordingly. In particular, in \cite{challa} a buckling of these "pseudolayers" under the action of an external magnetic field was observed until they flatten for sufficiently high fields. There, a Helfrich-Hurault model was proposed to theoretically describe this phenomenon and infere the value of the associated elastic constants from the experimental values of the critical magnetic fields. On the other hand, the $N_{TB}$ configurations presented here, and derived from the phenomenological elastic theory \eqref{virga_free_energy2_mag}, uniformly fill the space with heliconical axes parallel to each other, meaning that the surfaces of constant azimuthal phase are flat independently of the presence of an external field.
In order to allow for displacements from flat surfaces, one should add extra energy terms to \eqref{virga_free_energy2_mag} as reported in \cite{stewie2003} (see eq.(61) therein) by considering a suitable parametrization of $\bm{n}$ in terms of the unit normal to the surface displacement and the unit orthogonal projection of $\bm{n}$ onto it.

At this stage, additional comments on the constant pitch in the presence of external fields are in place.
In \cite{lavre2014} the effects of external fields on the tilt angle and the pitch of oblique helicoidal configurations in doped CB7CB were analyzed. The experimental results were in good agreement with the predictions of a theoretical model, first introduced in \cite{meyer1968}, based on the chiral Frank-Oseen framework. It turned out that both the pitch and the tilt angle increase as the field decreases from the critical value for the cholesteric-isotropic transition, originating a new configuration called oblique helicoid. Moreover, the CB7CB is known to be one of the recently discovered materials exhibiting the $N_{TB}$ phase. However, it is difficult to compare the results presented here with those obtained in \cite{lavre2014} since  quite different situations are examined. Indeed, the oblique $N^*$ phase, obtained through the addition of the chiral dopant,  shows a period of some microns, while, if no chiral molecules are added from outside, the periodic structures typical of the $N_{TB}$ phase have periods of a few nanometers. Furthermore, because of the dopant additive, the oblique helicoidal phase has the same chirality everywhere, whilst the $N_{TB}$ phase is formed by both left- and right-handed domains. Finally, by fitting theoretical predictions to experimental results, the authors in \cite{lavre2014} found the bend constant to be significantly smaller than others but still positive. We must stress that here we investigate an achiral scenario in which this latter constant turns negative, possibly allowing for the stabilization of $N_{TB}$ and other non-uniform configurations. Indeed, in \cite{lavre2014} it is stated that the oblique helicoids appear in a specific range of intensity of the external field, which allows for the chiral twist to compete with the torque of the field. Above this range, the homeotropic alignment is favored; below this range, the right-angle helix typical of cholesterics is stabilized. The limiting values for the amplitudes of the external field are both determined by the ratio $0<\kappa=K_{33}/K_{22}<1$: a negative value of $K_{33}$ would imply imaginary values for all the observables taken into consideration in the theoretical model proposed in \cite{lavre2014}.
Thus, it is not surprising that in our case the pitch keeps independent of the external field.

\section{Skyrmion tubes under external fields}
\label{sectfour}

\subsection{Skyrmion tube parameterization}

At variance with the previous section, here we consider the case of nonuniform distortions leading to localized states.
Bearing in mind that the uniform distortions are heliconical states, we slightly depart from this case by considering still heliconical structures,
but with a nonuniform conical angle and an additional precession around the heliconical axis.
These structures give rise to localized cylindrically-symmetric configurations, which can be referred to as {\emph{ Skyrmion tubes}} (SkT), of the general form
\begin{equation}
\label{SKT_ansatz}
\bm{n}( r, z,\varphi;\beta)=\sin(f(r))\cos( \varphi+\beta z)\bm{e}_x+\sin(f(r))\sin( \varphi+\beta z)\bm{e}_y+\cos(f(r))\bm{e}_z.
\end{equation}
The $\varphi-$dependence prescribes a winding performed by the director around the heliconical axis $\bm{e}_z$ for fixed $z$,
$f(r)$ is the profile function describing the conical angle and
$\beta$ has the same meaning as in the previous section.

One might also think about an ansatz without the angular dependence given by $\varphi$.
However, as we already discussed in \cite{pre102us} and also checked from numerical calculations and within this setup under an external field, stable solutions of this kind do not exist. Hence, this ansatz with no $\varphi-$dependence gives us the uniform distortion as the ground state where now, the conical angle $\theta_0$ will also depend on the value of the external field as showed in the previous section.
The winding around the $z$ axis given by the azimuthal variable is of key importance to give rise to these localized Skyrmion tubes and prevent them from directly decaying into the ground state.

In order to justify the above ansatz, one way  to proceed is to resort to the so-called reduction by variational point symmetries \cite{olver}, where these latter transform both independent and dependent variables, leaving unchanged the value of the functional (\ref{free_energy_volume}) and
the associated full Euler-Lagrange equations (\ref{euler_lagrange_eq_virga_f}). 
A symmetry reduction procedure leads to an exact form of the solution with a less number of independent variables, as in (\ref{SKT_ansatz}), and it produces the corresponding equations in the remaining unknown functions. These latter obey to a restricted class of boundary conditions.
Finding all symmetries admitted by the equations (\ref{euler_lagrange_eq_virga_f})  may be challenging because of their high complexity. Still, one can exploit  the constructive assumptions of  translational and rotational invariance \cite{Virga4}.  To this purpose, it is useful to parametrize the director field $\bf n$ in terms of two real stereographic variables $\rho$ and $\Phi$ according to the following correspondence
\begin{equation}
w=\rho(x, y, z ) \exp \left(\imath \Phi\left( x, y, z \right)\right) \in \mathbb{C}\leftrightarrow \bm{n}=\frac{w+\bar{w}}{1+|w|^2}\bm{e}_x+\frac{-\imath(w-\bar{w})}{1+|w|^2}\bm{e}_y+\frac{1-|w|^2}{1+|w|^2}\bm{e}_z.
\end{equation}
By expressing the infinitesimal point symmetries in terms of vector-fields $\vec{v}$ in the space of the independent and dependent variables,
one can prove that any 1-dimensional subalgebra of the class
\begin{equation}
\vec{v} = \alpha \p_z + y \p_x - x \p_y - \p_{\Phi} \quad \alpha \in \mathbb{R}
\end{equation}
i) leaves invariant the external magnetic field, ii) is a variational symmetry and  iii) admits the following invariants:
\begin{equation}
I_1 = \rho, \quad I_2 = \Phi + \frac{1}{\alpha} z,
\end{equation}
\begin{equation}
 I_3 = x^2+y^2 = r^2 ,\quad  I_4 =\frac{1}{\alpha} z + \arctan (\frac{y}{x}) = \frac{1}{\alpha} z + \varphi.
\end{equation}

Thus, one can claim that the original variational problem has symmetry invariant solutions of the form 
\begin{equation}
\rho = \rho ( r, \zeta) , \quad   \Phi = F(r, \zeta) -  \frac{1}{\alpha} z , \quad \zeta = I_4,
\end{equation}
where $\rho$ and $F$ are  functions to be determined by a pair of  symmetry-reduced partial differential equations in the independent variables $r, \zeta$ only. Now, supposing that the function $F$ is smooth and non trivial in the angle $\varphi$, then it has to be independent of $r$ in order to avoid discontinuities. Similarly, a non trivial dependence on $r$ of $\rho$ implies independence of $\zeta$, otherwise it may lead to singularity and multi-valuedness  in $\varphi$. Furthermore, by choosing in particular $F= -\zeta$ and identifying $2/\alpha = \beta$, we get the ansatz (\ref{SKT_ansatz}). Once the ansatz is justified, one can place it directly in the functional and find the corresponding reduced Euler-Lagrange equation for $f(r)$, as detailed in the following.

In order to have localized configurations, we may impose the boundary conditions $f(0)=0$ and $f(r\to\infty)=\theta_0$, $\theta_0$ being a suitable conical angle to be determined.
We also consider the case $f(0)=\pi$ and $f(r\to\infty)=\theta_0$.
Then, to study these configurations, we need to reduce the general free--energy in order to translate the ansatz into the equilibrium equations.
The reduced free--energy integrated over the unit cell $\left[0,\frac{2\pi}{\beta}\right]\times\left[0,2\pi\right]$ and over $r\in\left[0,\infty\right]$ will take the form
\begin{equation}
\mathcal{F}_{H}\left[f;\beta\right]=\int_{0}^{\frac{2\pi}{\beta}}{\text{d}z}\int_{0}^{2\pi}{\text{d}\varphi}\int_{0}^{\infty}F_{H}\left[\bm{n}(r, z,\varphi;\beta)\right]r\text{d}r.
\end{equation}
We are interested in the reduced free--energy per unit cell $\left[0,\frac{2\pi}{\beta}\right]\times\left[0,2\pi\right]$
which can be obtained by dividing by the factors $2\pi$ and $\frac{2\pi}{\beta}$
\begin{equation}
\tilde{\mathcal{F}}_{H}\left[f;\beta\right]=\frac{\beta}{4\pi^2}\mathcal{F}_{H}\left[f;\beta\right].
\end{equation}
This latter can be rewritten as follows
\begin{equation}
\label{unique_fe}
\tilde{\mathcal{F}}_{H}\left[f;\beta\right] = \frac{1}{256}\int \left( G_{\rm H} + G_0 + G_1 f' + G_2 f'^2 + G_3 f'^3 + G_4 f'^4  \right) \text{d}r,
\end{equation}
where $G_i$ are reported in the Appendix \ref{app:equation} and we have now defined 
\begin{equation}
G_{\rm H} = - 128 \chi_a H^2 r \cos^2 f.
\end{equation}
Hence, we arrive at the following Euler-Lagrange associated equation
\begin{eqnarray}
\label{skyeq}
&& 2f^{\prime\prime}\left(G_2+3f^{\prime}G_3+6f^{\prime 2}G_4\right)+2f^{\prime}\partial_rG_2+f^{\prime 2}\partial_fG_2+2f^{\prime 3}\left(\partial_fG_3+2\partial_rG_4\right) \nonumber \\
&& +3f^{\prime 4}\partial_fG_4+\partial_rG_1-\partial_fG_0 - \partial_f G_{\rm H}=0.
\end{eqnarray}
As stated above, we will look for localized solutions of the form (\ref{SKT_ansatz}).
The radial profile function $f(r)$ solves the ODE (\ref{skyeq}). Here, we want to study the asymptotic behaviour as $r\to\infty$.
The asymptotic state will then be denoted as
\begin{equation}
\bm{n}_{\infty}=\sin \theta_0\cos(\varphi+\beta z)\bm{e}_x+\sin \theta_0\sin(\varphi+\beta z)\bm{e}_y+\cos \theta_0\bm{e}_z,
\end{equation}
where $\theta_0$ is the asymptotic conical angle, {\emph{i. e.}}, $f(r)\to \theta_0$ as $r\to \infty$.

In order to determine $\theta_0$, as in the case of uniform distortions, we follow the route of free--energy minimization.
Alternatively, we could study the asymptotic behaviour directly from (\ref{skyeq}).
To find this value, we just need to consider the stationary condition of the free--energy with respect to $f$. In this way, we get an asymptotic angle depending only on the elastic constants, the external field, and the $\beta$ parameter
given in terms of the elastic constants only (\ref{beta}).
The free--energy to be minimized is the asymptotic expression of $\tilde{\mathcal{F}}_{H}$ as in (\ref{unique_fe}).
As $r\to\infty$, we get
\begin{equation}
\tilde{\mathcal{F}}_{H}[f;\beta]= \int \left[ \frac{1}{256}G_0^{\infty} (r, f)-\frac{\chi_a}{2}H^2r\cos^2f  \right] \text{d}r + \text{h.o.t.} \, ,
\end{equation}
where the function $G_0^{\infty}$ is obtained from the function $G_0$ by dropping
all the terms $1/r$ and $1/r^3$ and keeping only linear terms in $r$, {\emph{i. e.},
\begin{eqnarray}
G_{0}^{\infty} = g_{01}^{\infty} + g_{02}^{\infty}\cos 2 f+ g_{03}^{\infty}\cos 4 f + g_{04}^{\infty}\cos 6 f + g_{05}^{\infty}\cos 8 f,
\end{eqnarray}
where
\begin{eqnarray}
g_{01}^{\infty}& =& \frac{\beta^2}{2}r(192k_2+32k_3+70\beta^2k_4+3\beta^2k_5-10\beta^2k_6),
\end{eqnarray}
\begin{eqnarray}
g_{02}^{\infty}&=&-4\beta^2 r(32k_2+14\beta^2k_4-k_6\beta^2),
\end{eqnarray}
\begin{eqnarray}
g_{03}^{\infty}&=&+2\beta^2 r(16k_2-8k_3+14\beta^2 k_4-\beta^2 k_5+2\beta^2 k_6),
\end{eqnarray}
\begin{eqnarray}
g_{04}^{\infty}&=&-4\beta^4 r (2k_4+k_6),
\end{eqnarray}
\begin{eqnarray}
g_{05}^{\infty}&=&\frac{\beta^4}{2}(2k_4+k_5+2k_6)r,
\end{eqnarray}
entailing that
\begin{equation}
g_{01}^{\infty}+g_{02}^{\infty}+g_{03}^{\infty}+g_{04}^{\infty}+g_{05}^{\infty} = 0.
\end{equation}
We then need to minimize the function
\begin{equation}
f_{H}(\theta_0) = \frac{1}{256}G_0^{\infty}(r,\theta_0)-\frac{\chi_a}{2}r\cos^2\theta_0,
\end{equation}
where we are now using the asymptotic value of $f$, {\emph{i. e.}}, $f\to\theta_0$ as $r\to\infty$.
Dropping the $r$ in the above expression, the function to be minimized is
\begin{eqnarray}
f_{H}(\theta_0)=\frac{1}{256}\left[g_{01}^\infty+g_{02}^\infty\cos(2\theta_0)+g_{03}^\infty\cos(4\theta_0)+g_{04}^\infty\cos(6\theta_0)+g_{05}^\infty\cos(8\theta_0)\right]
-\frac{\chi_a}{2}H^2\cos^2\theta_0.
\end{eqnarray}
Upon setting as above $t=\sin^2\theta_0$, we arrive at
\begin{eqnarray}
f_{H}(t)=\frac{1}{4}\left[-2k_6(1-t)t+2k_4t^2+k_5(1-t)^2\right]t^2\beta^4+\frac{1}{2}\left[2k_2t+k_3(1-t)\right]t\beta^2-\frac{1}{2}\chi_aH^2(1-t).
\end{eqnarray}
The corresponding stationary condition reads
\begin{equation}
\partial_t f_{H}(t)=0,
\end{equation}
that is
\begin{equation}
\beta^2\left(k_3+4k_2t-2k_3t\right)+\beta^4t\left[t\left(-3k_6+4k_4t+4k_6t\right)+k_5\left(1-3t+2t^2\right)\right]+H^2\chi_a=0,
\end{equation}
which reproduces the same equation as (\ref{eqtobestudied}).
Accordingly, we obtain for the asymptotic conical angle the same expression as in the uniform
heliconical configuration (\ref{thetazero_emf}) together with its dependence on the uniform external magnetic field.
In particular, it follows that the asymptotic angle $\theta_0$ vanishes when $H\geq H_{\text{cr}}$ (\ref{hcritical}) (see also Figs. \ref{theta0-k3bis} and \ref{theta0-k4bis}).

In the next section, we will look for localized solutions to (\ref{skyeq})
by numerically minimizing the free--energy with a gradient flow method.

\subsection{Numerical results}
\label{sectfour_B}

As mentioned above, we performed numerical simulations to minimize \eqref{free_energy_volume}  via the Euler-Lagrange equation (\ref{euler_lagrange_eq_virga_f}) in order to find the uniform heliconical state.
The same approach might be applied to find general configurations for a given direction of the external field and the boundary conditions. However, due to the issues highlighted in the previous section,
the study of localized solutions with 3D simulations is a challinging project, outside the scope of the present paper.
Here, we find the configurations corresponding to the Skyrmion tubes by minimizing the free--energy (\ref{unique_fe}) within the ansatz (\ref{SKT_ansatz}). Hence, the profile $f(r)$ can be numerically obtained by using a gradient flow method. For this purpose, we consider a 1-dimensional lattice $\mathcal{L}$ of 1000 points with a lattice spacing $\Delta r = 0.02$, with spatial derivatives approximated by a fourth-order finite difference. Regarding the boundary conditions, we will consider two different cases concerning the value at the origin: $f(0) = 0$ and $f(0) = \pi$, where $f(r \rightarrow \infty) = \theta_0$. Although we know that solutions taking the value $\pi$ at the origin have higher energy \cite{pre102us}, they are interesting when placed under an external field since they may survive the application of a magnetic field bigger than $H_{\rm cr}$ (\ref{hcritical}). This should not be surprising since $\theta_0$ is a function not only of $\beta$ and the elastic constants but also of the external field $H$. Indeed, as an increasing $H$ will decrease the conical angle $\theta_0$, the configuration with $f(0) = 0$ will converge to the nematic phase when $H=H_{\rm cr}$, whereas the solution with $f(0) =\pi$ will remain, interpolating between $\pi$ at the origin and zero at infinity. In other words, we can say that when increasing the external field, the first class of solutions start to dilute in the ground state, {\emph{i. e.}}, the difference between the profile values at the origin and at infinite decreases until vanishing, leading to the uniform ground state $\bm{n}=\bm{e}_z$ everywhere. 

On the other hand, as for the second class, when $H\geq H_{\rm{cr}}$ the director assumes the two degenerate ground state configurations $\bm{n}=-\bm{e}_z$ and $\bm{n} = \bm{e}_z$ at $r=0$ and as $r\to \infty$, respectively. Thus, a non-uniform configuration around the center survives and a cylindrical domain wall connecting the two different ground states arises.

In Fig. \ref{Sol-Standard}, solutions with elastic constants $k_1 = k_2 = k_4 = k_5 = k_6 = 1$ and $k_3 = -3$ (the standard set) for different values of the external field are shown for these two different boundary conditions at the origin. For $\beta$ we have chosen its value in terms of the elastic constants as in (\ref{beta}), {\emph{i. e.}}, $\beta = 5$ in this case. All this gives us a critical field $H_{\rm cr}^2 = 75/\chi_a$. One can easily see the different effect of increasing the external field in each class of configurations due to the diminution of the conical angle $\theta_0$ with it. Moreover, Fig. \ref{Energies-Standard} depicts the energy per pitch, $P=\frac{2 \pi}{|\beta|}$, of both classes of solutions,
\begin{equation}
E_{\rm SkT} = \int_{0}^{\frac{2\pi}{\beta}}{\text{d}z}\int_{0}^{2\pi}{\text{d}\varphi}\int_{\mathcal{L}}F_{H}\left[\bm{n}(r, z,\varphi;\beta)\right]r\text{d}r,
\end{equation}
once the energy of the ground state, {\emph{i. e.}},
\begin{equation}
E_{\rm GS} = \int_{0}^{\frac{2\pi}{\beta}}{\text{d}z}\int_{0}^{2\pi}{\text{d}\varphi}\int_{\mathcal{L}}F_{H}\left[\bm{n}_h\right]r\text{d}r,
\end{equation}
is subtracted, {\it i. e.}, $\Delta E /P = (E_{\rm SkT} - E_{\rm GS})/P$ (we have introduced the notation $E$ for the energy coming from the numerical calculation to make clearer it is calculated in the finite lattice $\mathcal{L}$). For the configurations with $f(0)=0$, the excitation energy decreases when approaching the critical field, as expected since for $H=H_{\rm cr}$ they converge to the nematic phase. Otherwise, if $f(0)=\pi$ the excitation energy remains about the same value for small fields before rapidly increasing.
\begin{figure}[ht]
\begin{center}
\includegraphics[width=.75\textwidth]{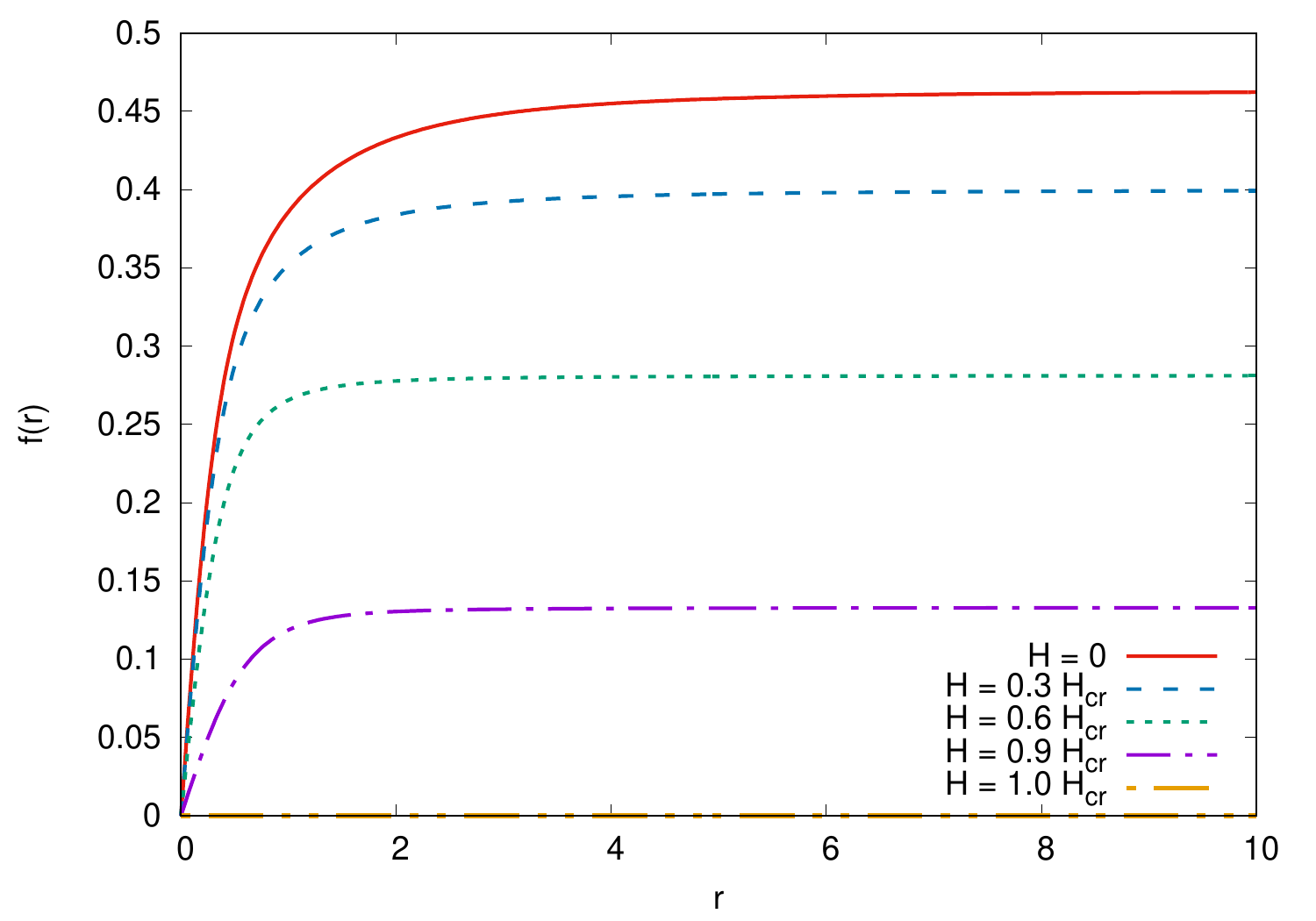}
\includegraphics[width=.75\textwidth]{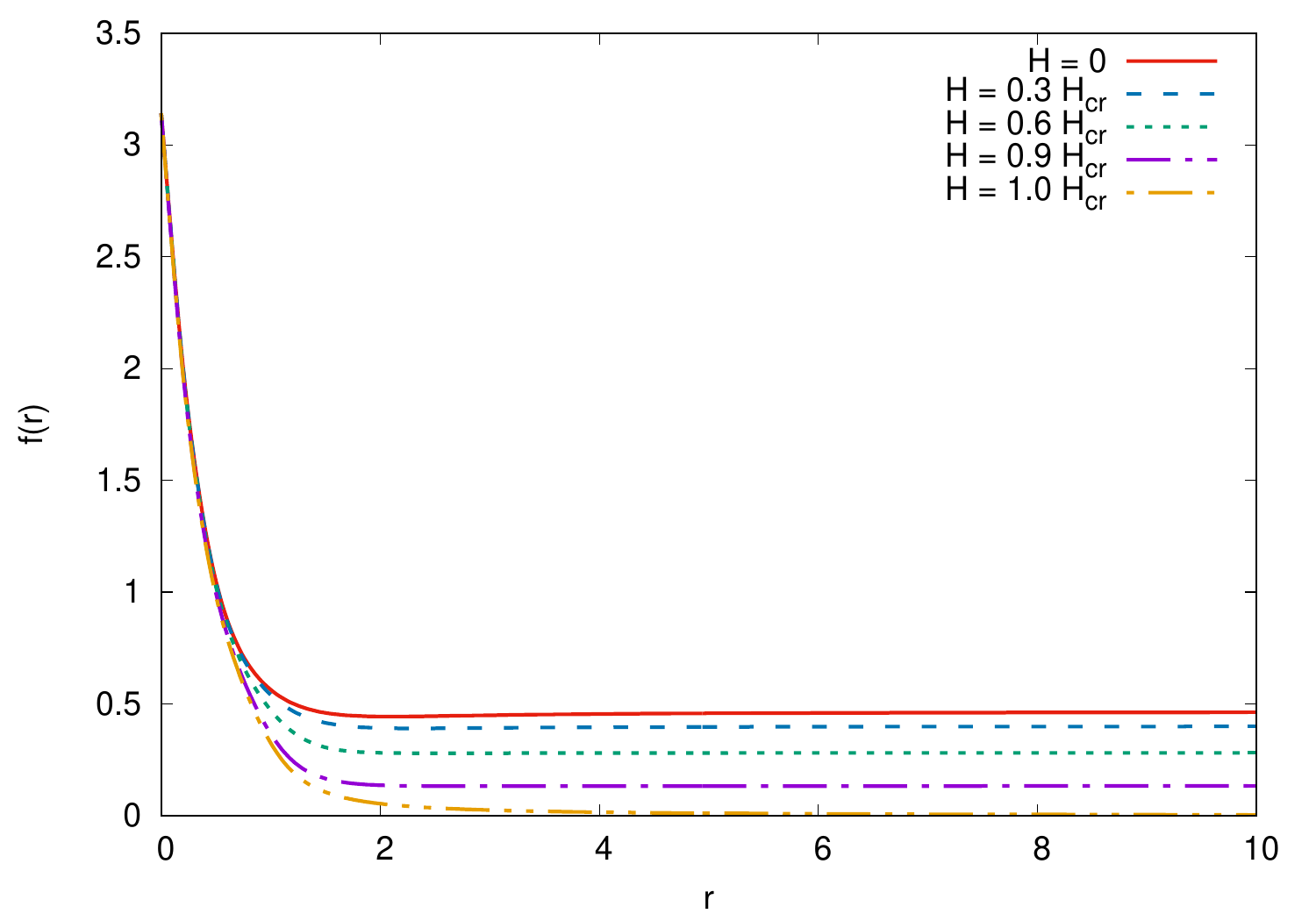}
\end{center}
\caption{Profile solutions of the conical angle for the elastic constants $k_1 = k_2 = k_4 = k_5 = k_6 = 1$ and $k_3 = -3$, both when $f(0)=0$ (up) and $f(0)=\pi$ (down). For these values of the elastic constants, $\beta = 5.0$ and $H_{\rm cr}^2 = 75/\chi_a$.}
\label{Sol-Standard}
\end{figure} 
\begin{figure}[ht]
\begin{center}
\includegraphics[width=.75\textwidth]{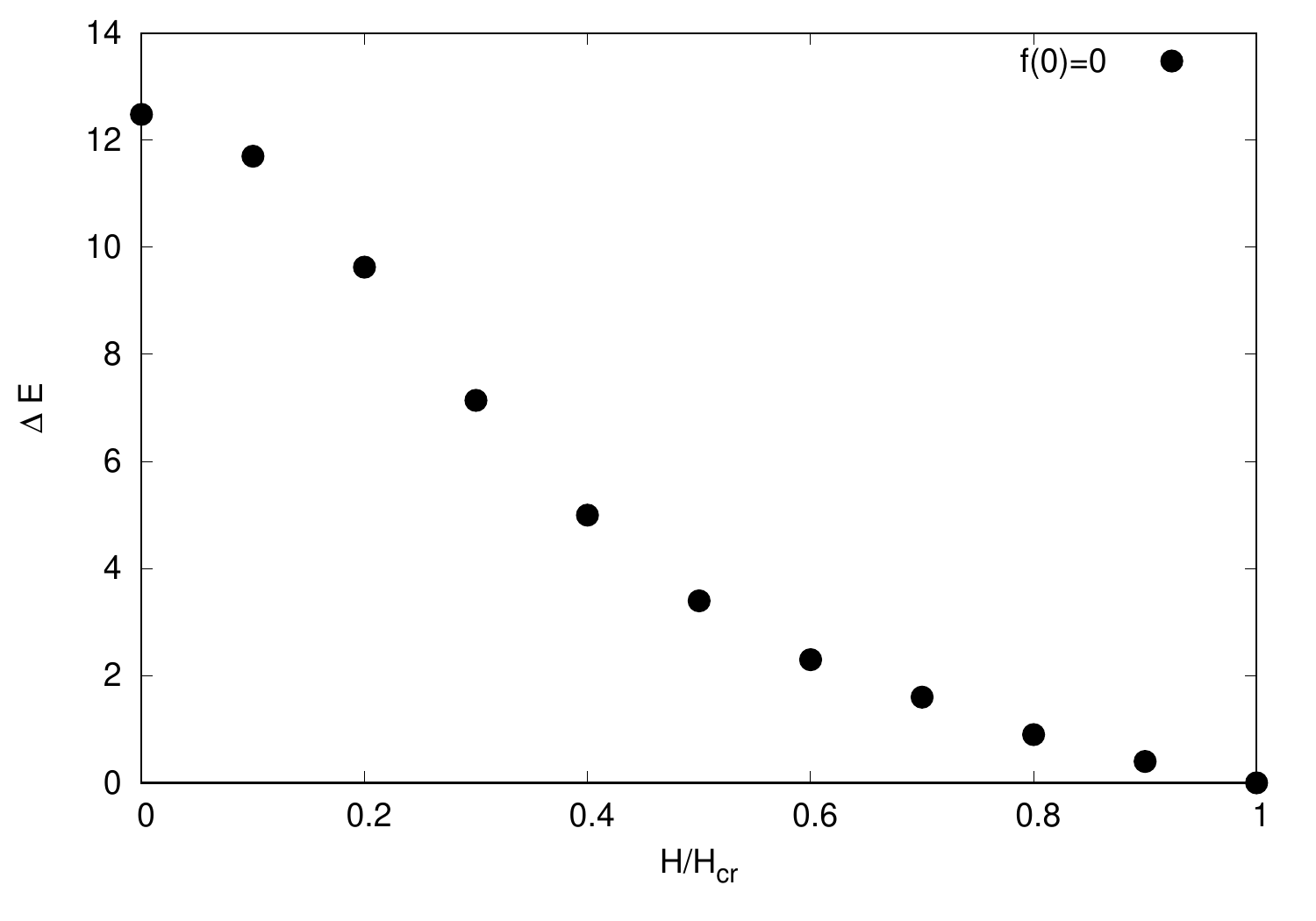}
\includegraphics[width=.75\textwidth]{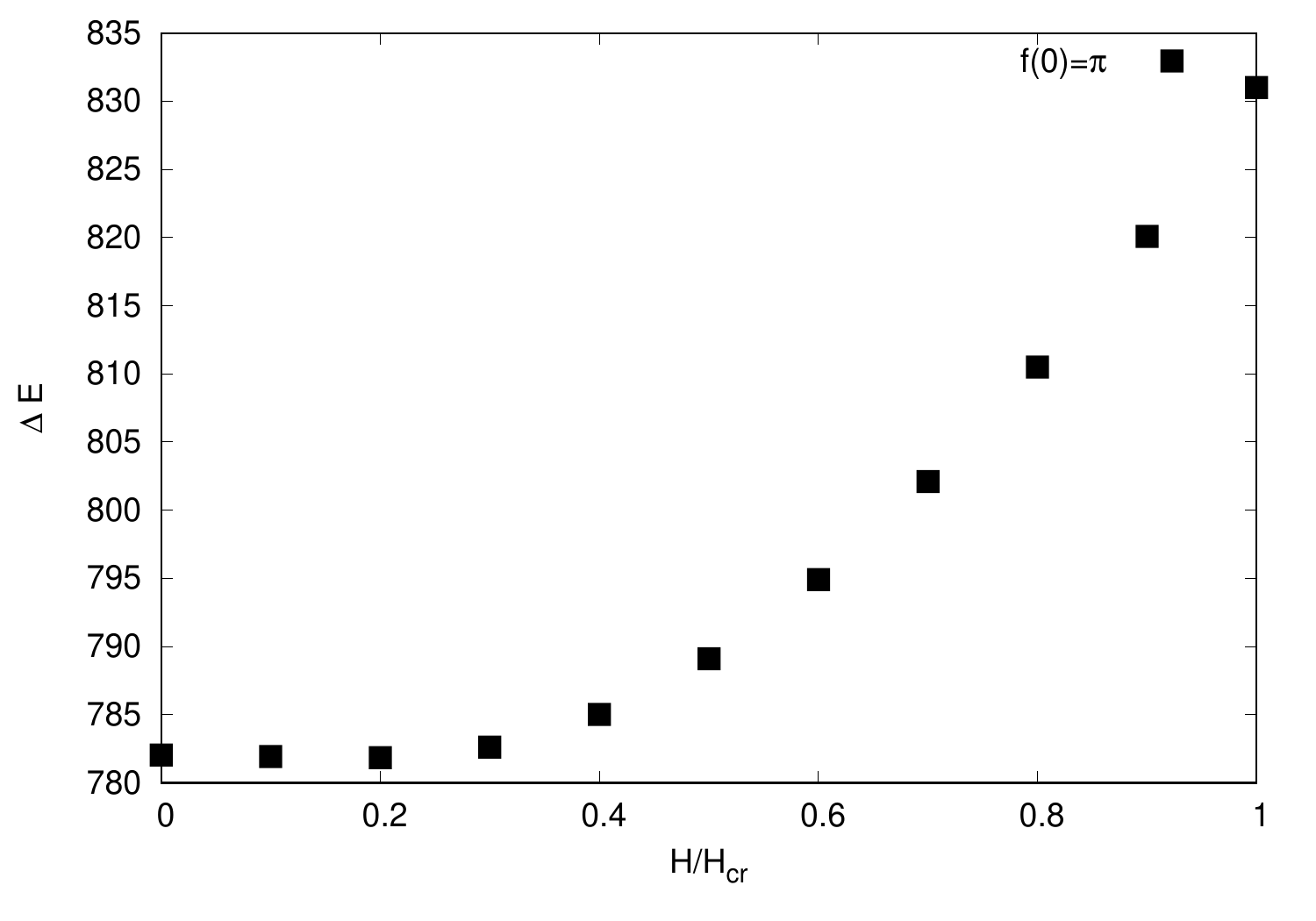}
\end{center}
\caption{Excitation energy with respect to the ground state as a function of the external field for the solutions with elastic constants $k_1 = k_2 = k_4 = k_5 = k_6 = 1$ and $k_3 = -3$, both when $f(0)=0$ (up) and $f(0)=\pi$ (down). For these values of the elastic constants, $\beta = 5.0$ and $H_{\rm cr}^2 = 75/\chi_a$.}
\label{Energies-Standard}
\end{figure} 
One can also go further and explore a little bit the parameter space. For instance, we can decrease the elastic constant $k_3$. Doing this, the behaviour of the profile $f(r)$ is similar to the one shown in Fig. \ref{Sol-Standard}, since an increasing external field always implies a diminution of the conical angle $\theta_0$ (see Figs. \ref{theta0-k3bis} and \ref{theta0-k4bis}).

Finally, one can consider how the excitation energy changes both with the external field and the elastic constants when $f(0) = \pi$. This has been depicted in Fig. \ref{Energies_fcPI} for different values of the coupling constants $k_3$ and $k_4$ in a logarithm scale. In this way, it is manifest how an increasing $k_4$ has an important effect, raising the excitation energies in a considerable way, much more remarkable than when the value of the elastic constant $k_3$ is varied with respect to the standard set of values. This may be due to the fact that, even in the absence of external field, the twist-bend phase is characterized by a smaller conical angle than in the standard case, making the central value $f(0) = \pi$ present a higher deviation from the ground state.

In order to obtain a better visualization of the nematic texture, in Figs. \ref{3dtube_nice} and \ref{tubecut_nice} we reported a three-dimensional representation of a Skyrmion tube with a profile function taking the value $\pi$ at the center, {\emph{i. e.}}, $f(0) = \pi$.
As it is clear, a tube is defined as an axially-symmetric region where the conical angle changes from $\pi$ to the asymptotic value $\theta_0$. When the external field increases, the tube is surrounded by a uniform nematic phase as $\theta_0$ vanishes
when $H\geq H_{\text{cr}}$.  Actually, in Fig. \ref{zoomprofile_nice} we also report the profile function when $H=H_{\text{cr}}, 1.5H_{\text{cr}}, 2H_{\text{cr}}$, where the shrinking of the Skyrmion tube with an increasing field is manifest.
\begin{figure}[ht]
\begin{center}
\includegraphics[width=.75\textwidth]{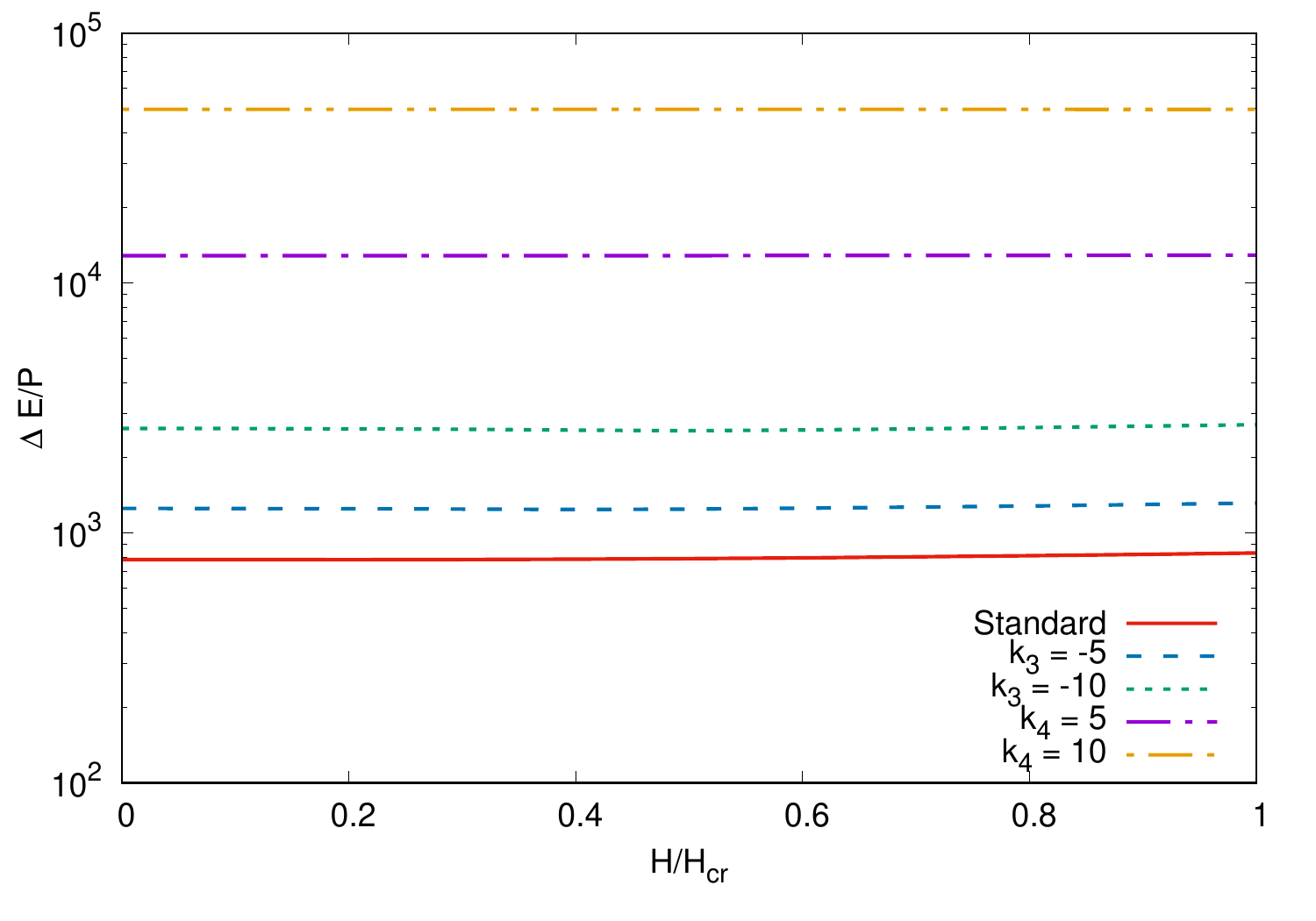}
\end{center}
\caption{Excitation energies with respect to the ground state as a function of the external field for different values of the coupling constants in logarithmic scale. The excitation energies correspond to the class of solutions with a profile function taking the value $\pi$ at the center, i.e., $f(0) = \pi$.}
\label{Energies_fcPI}
\end{figure} 
\begin{figure}[ht]
\begin{center}
\includegraphics[width=.75\textwidth]{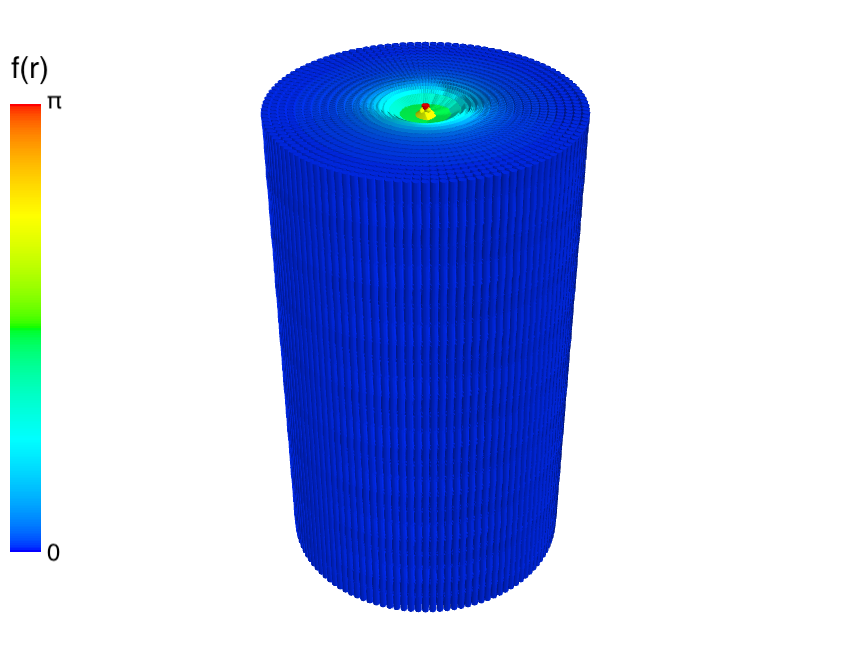}
\end{center}
\caption{Three-dimensional representation of a Skyrmion tube with a profile function taking the value $\pi$ at the center, {\emph{i.e.}}, $f(0) = \pi$. The cylinders show the vector director with the coloring indicating the conical angle.  }
\label{3dtube_nice}
\end{figure} 
\begin{figure}[ht]
\begin{center}
\includegraphics[width=.75\textwidth]{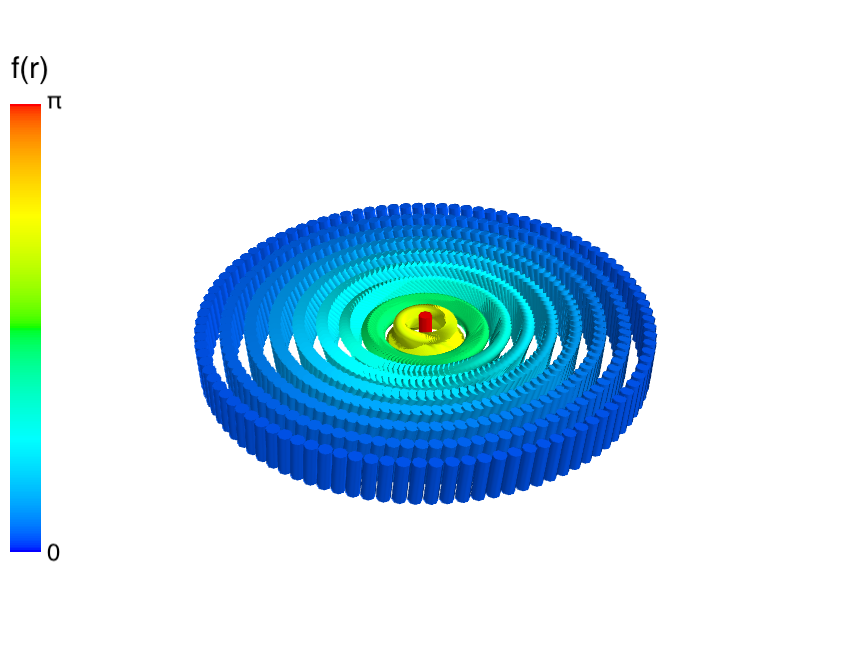}
\end{center}
\caption{A section of the three-dimensional representation of a Skyrmion tube with a profile function taking the value $\pi$ at the center, {\emph{i.e.}}, $f(0) = \pi$. The cylinders show the vector director with the coloring indicating the conical angle.}
\label{tubecut_nice}
\end{figure} 
\begin{figure}[ht]
\begin{center}
\includegraphics[width=.75\textwidth]{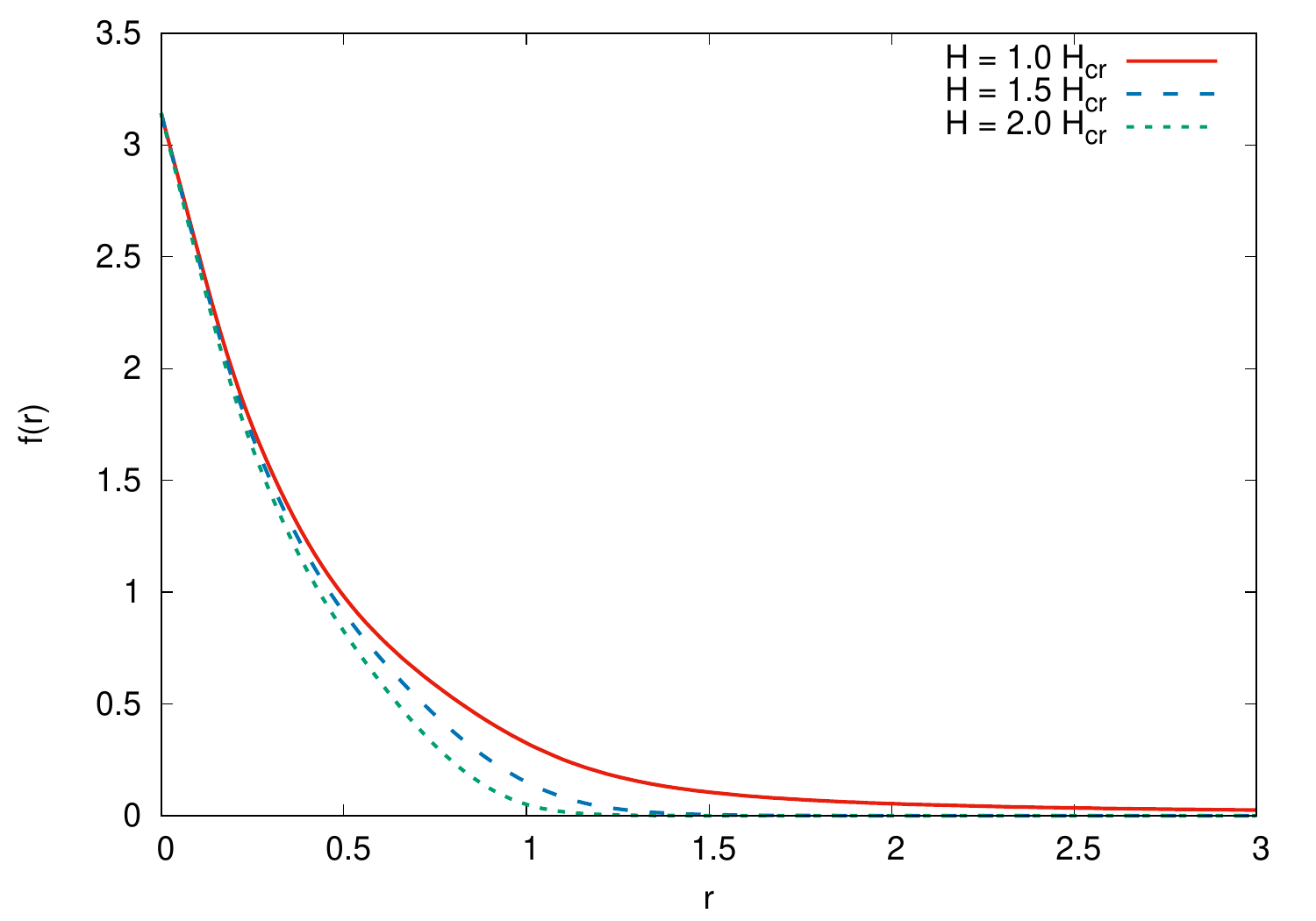}
\end{center}
\caption{Profile solutions of the conical angle for the elastic constants $k_1 = k_2 = k_4 = k_5 = k_6 = 1$ and $k_3 = -3$, when $f(0)=\pi$. For these values of the elastic constants, $\beta = 5.0$ and $H_{\rm cr} = 75.0$. The applied external magnetic field
is higher than the critical field $H_{\text{cr}}$.}
\label{zoomprofile_nice}
\end{figure} 

As for stability, we have some numerical evidence of it by energy considerations, although there is not  a complete proof based on the second variation of the free-energy functional.
Nevertheless, the calculation and analysis of the  second variation are not trivial, even for the quadratic Frank's functional \cite{stability_V}. 
This problem would deserve a separate further treatment and consideration, which are certainly beyond the scope of the present paper.

As a conclusion, the numerical analysis shows the existence of two families of solutions to equation \eqref{skyeq}.
In the Appendix \ref{app: ga_approx}, following an approach outlined in \cite{pre102us}, one may find a global analytical approximation which can fit these numerical solutions.

\section{Conclusions and perspectives}
\label{sectfive}

In this paper we studied the interaction of external uniform magnetic fields with achiral liquid crystals according to a generalized fourth-order
elasticity theory recently put forward in \cite{Virga4}. This theory is encoded in the free--energy density (\ref{virga_free_energy2}) which is parameterized by six
elastic constants: $k_1,k_2,k_3$ associated with the quadratic terms in the Frank free--energy and $k_4,k_5,k_6$ related to fourth-order contributions.
Under appropriate constraints on the six elastic constants, the proposed free--energy admits heliconical configurations as global minimizers.
They are characterized by 
a director forming a constant conical angle $\theta_0$ with respect to a fixed axis, say $z$, as shown in Fig. \ref{3D-alhpa0},
continuously precessing when moving parallel to this axis and turning completely round over the length of a pitch $P=2\pi/\beta$ (\ref{heliconics_1}).
These heliconical configurations have been recently identified experimentally in the ground state of twist-bend nematic phase $N_{\text{TB}}$.

When an external magnetic field is applied, an interaction term $\Gamma_{\text{H}}$ (\ref{virga_free_energy2_mag}) is added to the free--energy density. We studied
the effect of a uniform field along the symmetry axis of the uniform heliconical state. The heliconical uniform state preserves its pattern and
only the conical angle is affected by the external field as a consequence of the magnetic torque imparted to the nematic director.
As the magnitude of the magnetic field is increased, the nematic undergoes a transition from the $N_{\text{TB}}$ phase to the uniform nematic phase $N$
where the director lines up with the external field. The transition takes place at a critical value $H_{\text{cr}}$ (\ref{hcritical}).
The twist-bend phase has been further reproduced by 3D simulations through a minimization of the energy functional (\ref{free_energy_volume}) via Euler-Lagrange equations (\ref{euler_lagrange_eq_virga_f}). According to this, we should stress that the pitch remains constant, thus consistently confirming our initial assumption.

Following our previous work \cite{pre102us}, now in the presence of the external field, we generalized the heliconical configurations to nonuniform localized axially symmetric structures with a variable
conical angle (\ref{SKT_ansatz}).
Actually in \cite{pre102us}, in the zero-field case, we had shown that there exists an axially symmetric state where the conical angle
depends on the radial distance from the symmetry axis, the $z$-axis in our parameterization, going from $0$ (or $\pi$) to a $\theta_0$ at infinity in the radial direction, while
the director winds uniformly once around the $z$-axis. The conical angle profile goes from $0$ to its asymptotic value $\theta_0$ in an exponentially fast way, thus singling out
a central core. These localized structures are usually referred to as Skyrmion tubes \cite{Rybakov2015,Du2018,PRB98, PRB100}.
The free--energy corresponding to the configuration starting from $0$ at the origin of the radial axis has lower energy with respect to
the one starting from $\pi$, although they both are excited states with respect to the uniform heliconical distortion.
As shown in the present paper, for sufficiently low applied external fields, the Skyrmion tubes still keep their basic structure as localized configurations with a central core surrounded by a uniform
heliconical distortion and a winding around the symmetry axis. Once a critical threshold $H_{\text{cr}}$ (the same as in the uniform heliconical phase) is reached, the Skyrmion tubes undergo
a change in their patterns according to the value of the conical profile function at the origin of the symmetry axis.
More precisely, when the conical angle is zero at the origin and $H\geq H_{\text{cr}}$, the conical profile function vanishes
and the liquid crystal undergoes the transition to the uniform standard nematic phase, where the director lines up everywhere with the external field.
Thus, in this case the central core of the Skyrmion tube tends to disappear.
On the other hand, when the conical angle takes the value $\pi$ at the origin, at sufficiently high external fields the central core still survives
and it gets surrounded by a standard uniform nematic pattern where the nematic director lines up with the external field.
Thus, in the central core the conical angle changes rapidly from the value at the center to zero.    

As stated above, we would like to stress here that the configurations found in the present paper are of the same type as the so-called Skyrmion tubes found in \cite{PRB98, PRB100} and there described numerically
in ferromagnets and experimentally detected for chiral nematic liquid crystals under an applied external field. In contrast with these results, we found
Skyrmion tube configurations in achiral nematics either with external fields or in their absence. When a sufficiently high external field is present, only a type of Skyrmion tube survives
and it gives rise to an axially-symmetric localized configuration immersed in the standard nematic phase. 

Thus, we reached a twofold target. On one hand this work presents a self-contained study about the formation and control of Skyrmion tubes under external fields and, more specifically, coaxial external fields. On the other hand it represents a first stone towards a general 3D study of these structures also including arbitrary orientations of the external field.

As a conclusion, the proposal of considering higher order free--energy expansions, as opposed to higher derivative ones \cite{Dozov2001}, leads to interesting new perspectives in the liquid crystal science with many potential
technological applications, where Skyrmion tubes might play an important role. On the theoretical side, according to our results it is clear that this kind of configurations
emerge in a natural and straight way from the proposed energy and can be controlled by external fields.  

As for future work, we plan to study the stability of the found solutions, the mutual interaction of Skyrmion tubes, the space arrangement of two or more of them  and their lattice configurations. Furthermore, we aim at
studying electro-optical effects and exploring other types
of localized objects.
Moreover, we also aim at exploring the effect of the compression of pseudolayers by using an appropriate compression energy and a representation of the director in terms
of the geometric objects defining the layer surfaces as discussed at the end of Sec. \ref{secttwo}.
Finally, we are also interested in studying liquid crystals confined within specific geometries and modeled by the quartic
free--energy density along with its coupling with external fields.

\section{Acknowledgments}   
GDM is supported by the Dipartimento di Matematica e Fisica "E. De Giorgi", University of Salento through the grant {\it Studio analitico di configurazioni spazialmente localizzate in materia condensata e materia nucleare}.
LM has been partially supported by  INFN IS-MMNLP.
CN has been supported by the INFN grant  19292/2017 (MMNLP) {\it Integrable Models and Their Applications to Classical and Quantum Problems} and by the Olle Engkvist foundation, Grant No 204-0185. VT is partially supported
by he Ministry for Education, University and Research - MIUR, Italy.  

\appendix

\section{Mathematical details}
\label{app:equation}

In this Appendix we collect the basic main functions and coefficients appearing in the equilibrium equations for Skyrmion tubes.

The quantities $G_i$, $i = 0, 1, 2, 3, 4$ appearing in equation (\ref{skyeq}) depend on $r, f, \beta, k_1, k_2, k_3, k_4, k_5, k_6$ and are listed below:
\begin{equation}
G_0 = G_0\left(r, f\right)=
g_{01}+g_{02}\cos(2f)+g_{03}\cos(4f)+g_{04}\cos(6f)+g_{05}\cos(8f) ,
\end{equation}
where
\begin{eqnarray}
g_{01}& =& \frac{1}{16r^3}(178k_4+105k_5-30k_6)+\frac{1}{2r}(64k_1+96k_2+48k_3+70\beta^2k_4+15\beta^2k_5-18\beta^2k_6)\nonumber\\
&+&\frac{\beta^2}{2}r(192k_2+32k_3+70\beta^2k_4+3\beta^2k_5-10\beta^2k_6),
\end{eqnarray}
\begin{eqnarray}
g_{02}&=&-\frac{1}{2r^3}(25k_4+21k_5-3k_6)-\frac{32}{r}(k_1+k_2+k_3)-\frac{2\beta^2}{r}(21k_4+3k_5-10k_6)\nonumber\\
&-&4\beta^2 r(32k_2+14\beta^2k_4-k_6\beta^2),
\end{eqnarray}
\begin{eqnarray}
g_{03}&=&\frac{1}{4r^3}(2k_4+21k_5+6k_6)-\frac{2}{r}(8k_2-4k_3+3\beta^2k_5+10\beta^2k_6)\nonumber\\
&+&2\beta^2 r(16k_2-8k_3+14\beta^2 k_4-\beta^2 k_5+2\beta^2 k_6),
\end{eqnarray}
\begin{eqnarray}
g_{04}&=&-\frac{1}{2r^3}(-k_4+3k_5+3k_6)+\frac{2\beta^2}{r}(5k_4+3k_5+6k_6)-4\beta^4 r (2k_4+k_6),
\end{eqnarray}
\begin{eqnarray}
g_{05}&=&(2k_4+k_5+2k_6)(\frac{3}{16r^3}-\frac{3\beta^2}{2r}+\frac{\beta^4}{2}r),
\end{eqnarray}
with
\begin{equation}
\label{id1}
g_{01}+g_{02}+g_{03}+g_{04}+g_{05}=0.
\end{equation}
As for $G_1$
\begin{eqnarray}
G_1=G_1(r,f)=g_{11}\sin{(2f)}+g_{12}\sin(4f)+g_{13}\sin(6f),
\end{eqnarray}
where
\begin{eqnarray}
g_{11} = \frac{1}{r^2}(-35k_4+5k_6)+64(k_1-k_2)-20(k_4+k_6)\beta^2,
\end{eqnarray}
\begin{eqnarray}
g_{12} = \frac{4}{r^2}(4k_4-k_6)+16(k_4+k_6)\beta^2,
\end{eqnarray}
\begin{eqnarray}
g_{13} = \frac{1}{r^2}(k_4+k_6)-4(k_4+k_6)\beta^2.
\end{eqnarray}
As for $G_2$
\begin{eqnarray}
G_2=G_2(r,f)=g_{21}+g_{22}\cos(2f)+g_{23}\cos(4f)+g_{24}\cos(6f),
\end{eqnarray}
where
\begin{eqnarray}
g_{21}=\frac{1}{r}(71k_4+5k_5+29k_6)+4r\left[8(k_1+5k_2+k_3)+\beta^2(37k_4+k_5+9k_6)\right],
\end{eqnarray}
\begin{eqnarray}
g_{22} = -\frac{1}{2r}(15k_4+15k_5+79k_6)+2r\left[16(k_1+k_2-k_3)-\beta^2(97k_4+k_5+17k_6)\right],
\end{eqnarray}
\begin{eqnarray}
g_{23}=\frac{1}{r}(-63k_4+3k_5+11k_6)+4\beta^2 r(11k_4-k_5-k_6),
\end{eqnarray}
\begin{eqnarray}
g_{24}=(k_4+k_5+k_6)(2\beta^2 r-\frac{1}{2r}).
\end{eqnarray}
The function $G_3$ is given by
\begin{equation}
G_3 = G_3(f) =g_{31}\sin(2f)+g_{32}\sin(4f),
\end{equation}
where
\begin{equation}
g_{31}=-8(6k_4+k_6),
\end{equation}
\begin{equation}
g_{32}=-4(4k_4-k_6).
\end{equation}
Finally,
\begin{equation}
G_4 = G_4(r,f)=g_{41}+g_{42}\cos(2f)+g_{43}\cos(4f),
\end{equation}
where
\begin{eqnarray}
g_{41} = r(65k_4+9k_5-8k_6),
\end{eqnarray}
\begin{eqnarray}
g_{42}=4r(5k_4-3k_5+2k_6),
\end{eqnarray}
\begin{eqnarray}
g_{43}=3r(k_4+k_5).
\end{eqnarray}
Notice that in addition to (\ref{id1}) other identities exist. This is due to the fact that there are seven free parameters: six independent elastic constants $k_i$ and a parameter $\beta$.

\section{Global approximation}
\label{app: ga_approx}

Following the numerical analysis performed in section \ref{sectfour_B}, in this Appendix, using an approach outlined  in \cite{pre102us}, we look for a global approximation which can fit the numerical solutions found.

As a first step, one can observe that a smooth function, although found numerically, can be locally approximated by inverse trigonometric functions of an auxiliary rational function $s(r)$, but this is not a priori obvious when such a function comes as a solution to a nonlinear ODE with singular coefficients and boundary conditions. Thus, in principle one should analyze the singularities of the general solution in the complex plane of the  independent variable $r$ by resorting to several appropriate methods
(see for example \cite{musette}). We assume here that the general solution to our equation is a meromorphic function except for an essential singularity at infinity. Thus, the general solution can be expanded in series of poles, these latter depending on the given boundary conditions. In this perspective, one may look for an approximated solution in the spirit of the Pad\'e expansion \cite{padel} which, for the sake of simplicity, we truncated at the fourth order. In principle, all coefficients involved into such an expansion could be determined, but the complicate structure of the equation would make this study quite difficult and not significant to our purposes. Thus, in the following we adopt a mixed strategy consisting in evaluating the unknown coefficients (as in eq. (\ref{f0fit}) below) directly from the numerical solution, mainly in order to verify the consistency of the above arguments.

\medskip
\subsubsection{Case $f(0)=0$}
In this case we assume that the approximant of $f(r)$ can be written as
\beq
\label{fa0}
f_a(r)=\pi-\arccos\lf -1+\frac{r^2}{2} s(r)\rg,
\eeq
with $s(r)$ a still unknown function supposed to be well defined and bounded all over the domain $r\in [0,\infty[$. It is straightforward to verify that, under these hypotheses, $f_a(0)=0$. Moreover, in order to fulfill also the boundary condition as $r\to\infty$, 
it is required that $s(r\to\infty)\propto \frac{1}{r^2}$.
Actually, letting $r\to\infty$ and setting $f_a(r\to\infty)=\theta_0$, we get
\beq
\theta_0=\pi-\arccos\lf -1+\frac{1}{2} c_0 \rg,
\eeq
that is
\beq
c_0=2\lf -\cos \theta_0 +1\rg=4\sin^2\frac{\theta_0}{2}.
\eeq
Now, following \cite{pre102us} we assume $s(r)$ as the rational function
\beq
\label{s0}
s(r)=c_0\frac{\overline{a}+r^2}{r^4+\overline{c} r^2+\overline{d}},
\eeq
where $c_0,\;\overline{a},\;\overline{c},\;\overline{d}$ are real constants to be determined. Thus, we can write $f_a(r)$ as
\beq
\label{f0fit}
f_a(r)=\pi-\arccos\lf-1+2r^2\sin^2\frac{\theta_0}{2}\frac{\overline{a}+r^2}{r^4+\overline{c} r^2+\overline{d}} \rg.
\eeq
We can now perform a best fitting procedure between \eqref{f0fit} and the actual numerical solutions analyzed in section \ref{sectfour_B}. The results of the least square procedure are represented in Fig. \ref{fig:fit0}, where the numerical solutions for the set of elastic constants $k_1= k_2=k_4=k_5=k_6=1, \; k_3=-3$ (the standard set) and $H/H_{\text{cr}}=0,\;0.3,\;0.6,\;0.9$ are interpolated by \eqref{f0fit}.
On the bottom of each subfigure, the function $f(r)-f_a(r)$ is reported as an estimation of the goodness of the fit.
The values of the optimal $\overline{a},\;\overline{b},\;\overline{c}$ for the cases taken into consideration are reported in Table \ref{tab:fit0}.
\begin{figure}
\centering
\subfigure[]{\includegraphics[width=.48\linewidth]{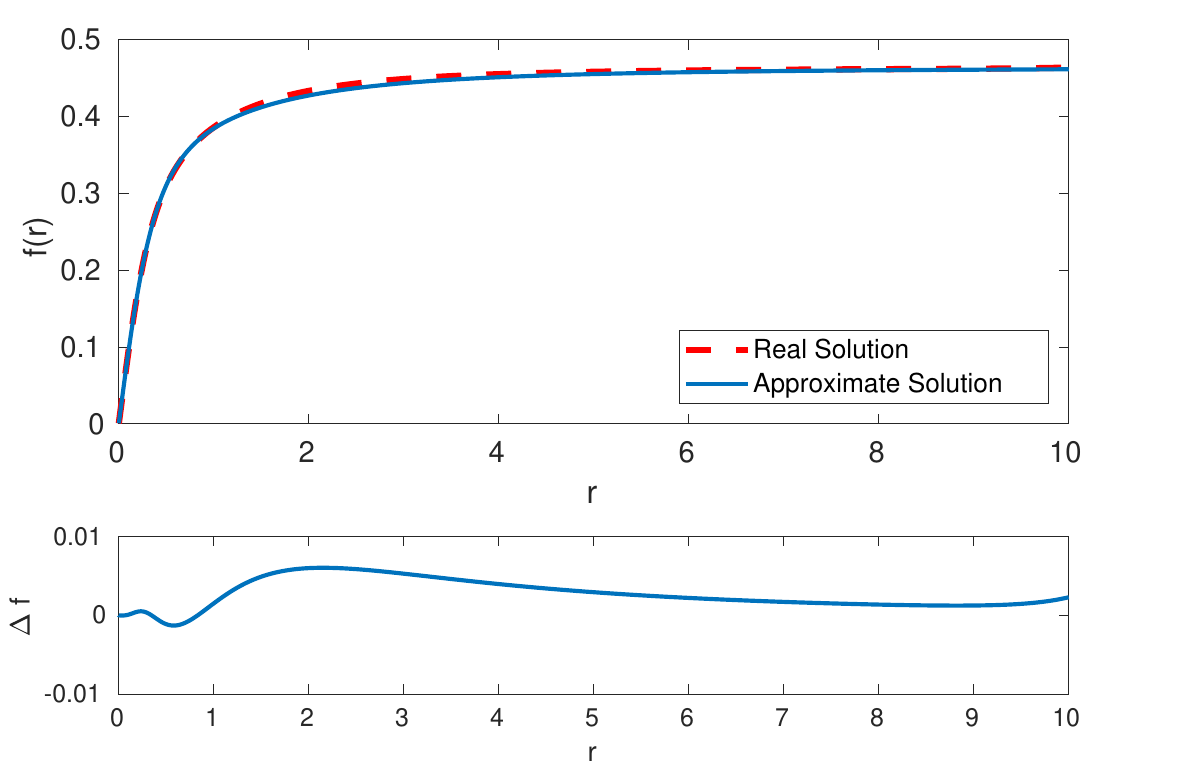}}  
\subfigure[]{\includegraphics[width=.48\linewidth]{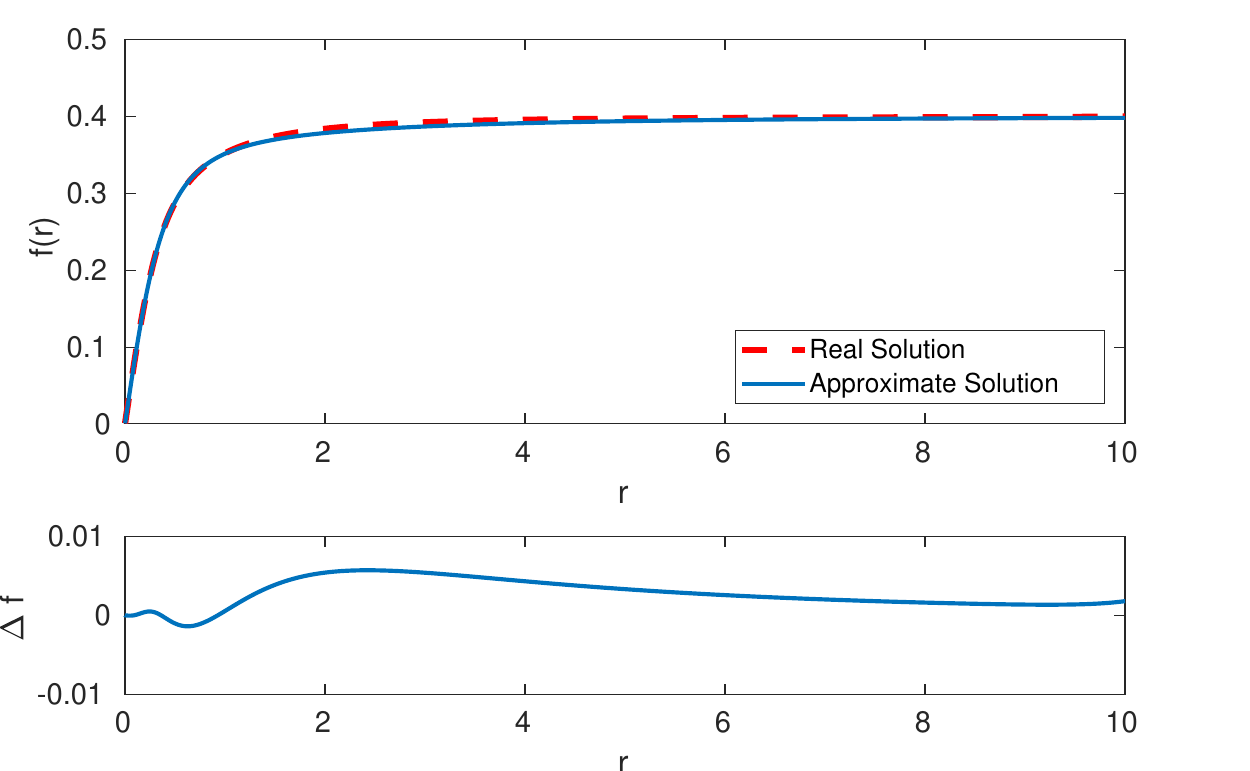}}
\subfigure[]{\includegraphics[width=.48\linewidth]{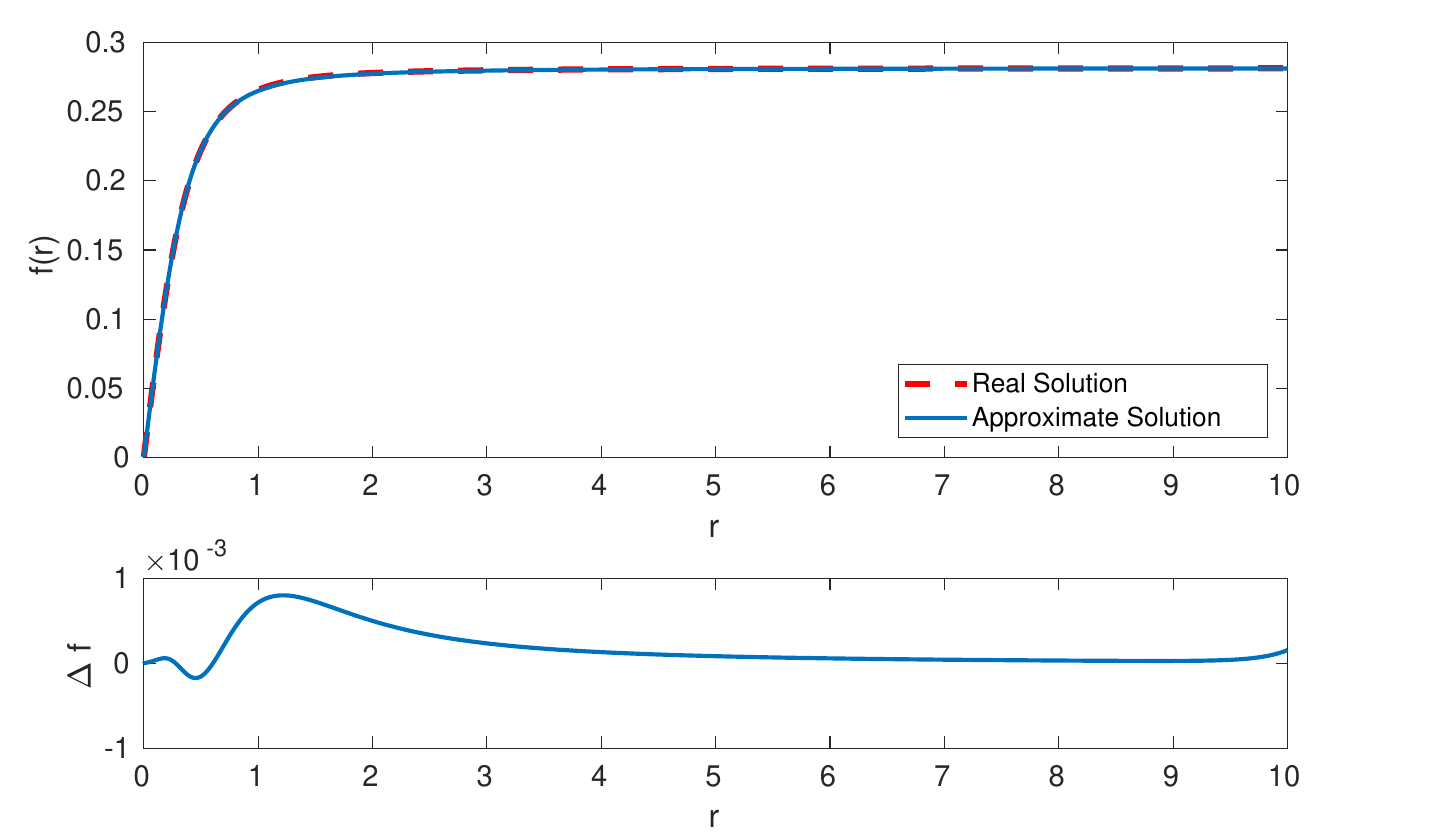}}
\subfigure[]{\includegraphics[width=.48\linewidth]{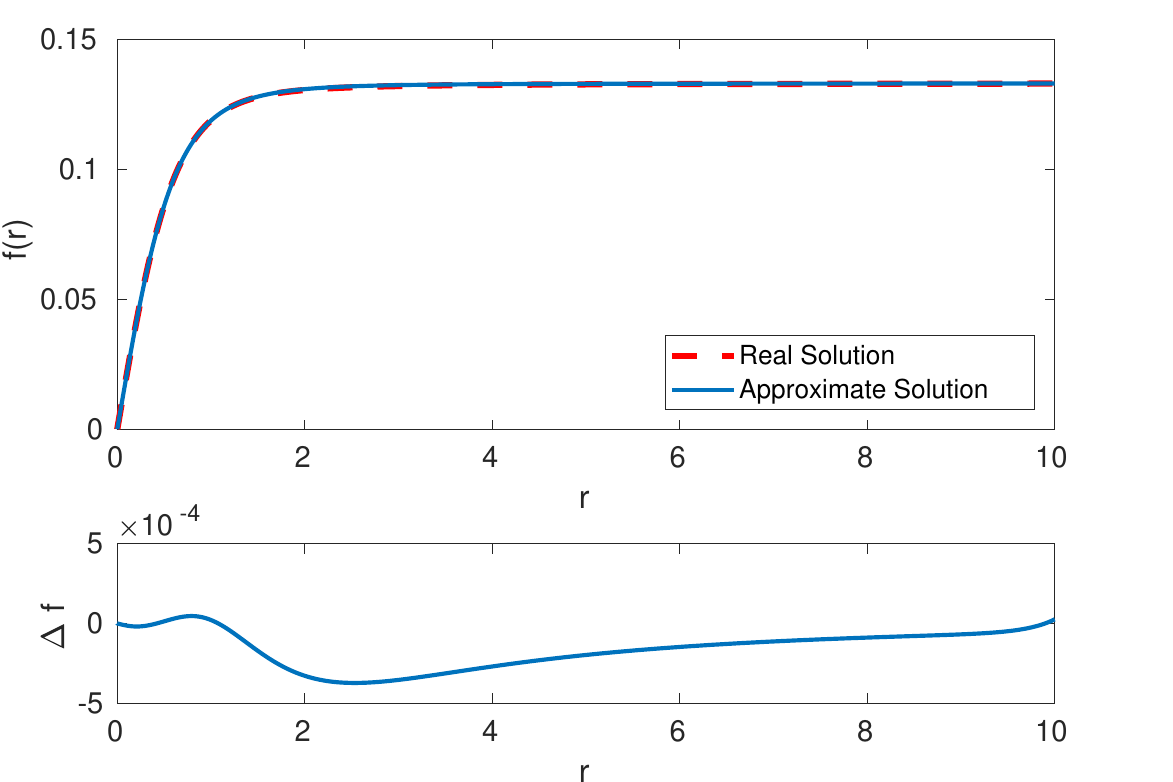}}
\caption{Best fitting results for $k_1= k_2=k_4=k_5=k_6=1, \; k_3=-3$ (the standard set) and $H/H_{\text{cr}}=\text{(a) }0,\;\text{(b) }0.3,\;\text{(c) }0.6,\;\text{(d) }0.9$ for the case $f(0)=0$. The red dashed line represents the numerical solution to equation \eqref{skyeq} and the blue one represents the best fitting curve \eqref{f0fit}. On the bottom of each subfigure, the function $f(r)-f_a(r)$ is reported as an estimation of the goodness of the fit.}
\label{fig:fit0}
\end{figure}

\begin{table}
\centering
\begin{tabular}{c|c|c|c}
$H/H_{\text{cr}}$ & $\overline{a} $ & $\overline{c}$ & $\overline{d}$ \\\hline
0				  & 2.6359  & 3.6417 & 0.6203 \\
0.3				  &  7.5163 & 8.4638 & 1.5279 \\
0.6				  &  0.3499 & 0.4586 & 0.0623 \\
0.9               & 1.5183  & 1.5514 & 0.6076 \\
\end{tabular}
\caption{Best fitting parameters for $k_1= k_2=k_4=k_5=k_6=1, \; k_3=-3$ and $H/H_{\text{cr}}=0,\;0.3,\;0.6,\;0.9$ for the case $f(0)=0$.}
\label{tab:fit0}
\end{table}
\medskip
\subsubsection{Case $f(0)=\pi$}
In this case, we write the approximant $f_a(r)$ as
\beq
\label{fpifit}
f_a(r)=\arccos\lf -1+\frac{r^2}{2} c_\pi \frac{\overline{a}+r^2}{r^4+\overline{c} r^2+\overline{d}}\rg.
\eeq
Similarly to the previous case, we find that $c_\pi$ reads
\beq
\label{cpi}
c_\pi=4\cos^2\lf\frac{\theta_0}{2} \rg,
\eeq
in order to fulfill the correct boundary conditions, \emph{i. e.}, $f_a(0)=\pi$ and $f_a(r\to\infty)=\theta_0$. 
Also in this case the best fitting procedure is successful, although for $H=H_{\text{cr}}$ the least square minimization must be performed for both the real and the imaginary part of $f_a$ simultaneously. More specifically, for every $r$ both quantities $\text{Re}[f(r)-f_a(r)]$ and $\text{Im}[f(r)-f_a(r)]$ are taken into account when minimizing the sum of the squared differences. Indeed, the minimization of only the real part  yields  values of the best fitting parameters such that $|-1+c_\pi\frac{r^2}{2} s(r)|>1$. As it can be noted also from the other cases, the ability of $f_a$ to fit $f$ is weaker in the proximity of the   bump through which the profile function reaches its asymptotic value. This is particularly true approaching the critical value of the external field.
Thus, the values of $\overline{a},\; \overline{c},\; \overline{d}$ in this latter case are the best fitting ones when a reality condition is imposed to $f_a$.
The results are depicted in Fig. \ref{fig:fitpi}. The values of the optimal $\overline{a},\;\overline{c},\;\overline{d}$ for the cases taken into consideration are reported in Table \ref{tab:fitpi}.
\begin{figure}
\centering
\subfigure[]{\includegraphics[width=.48\linewidth]{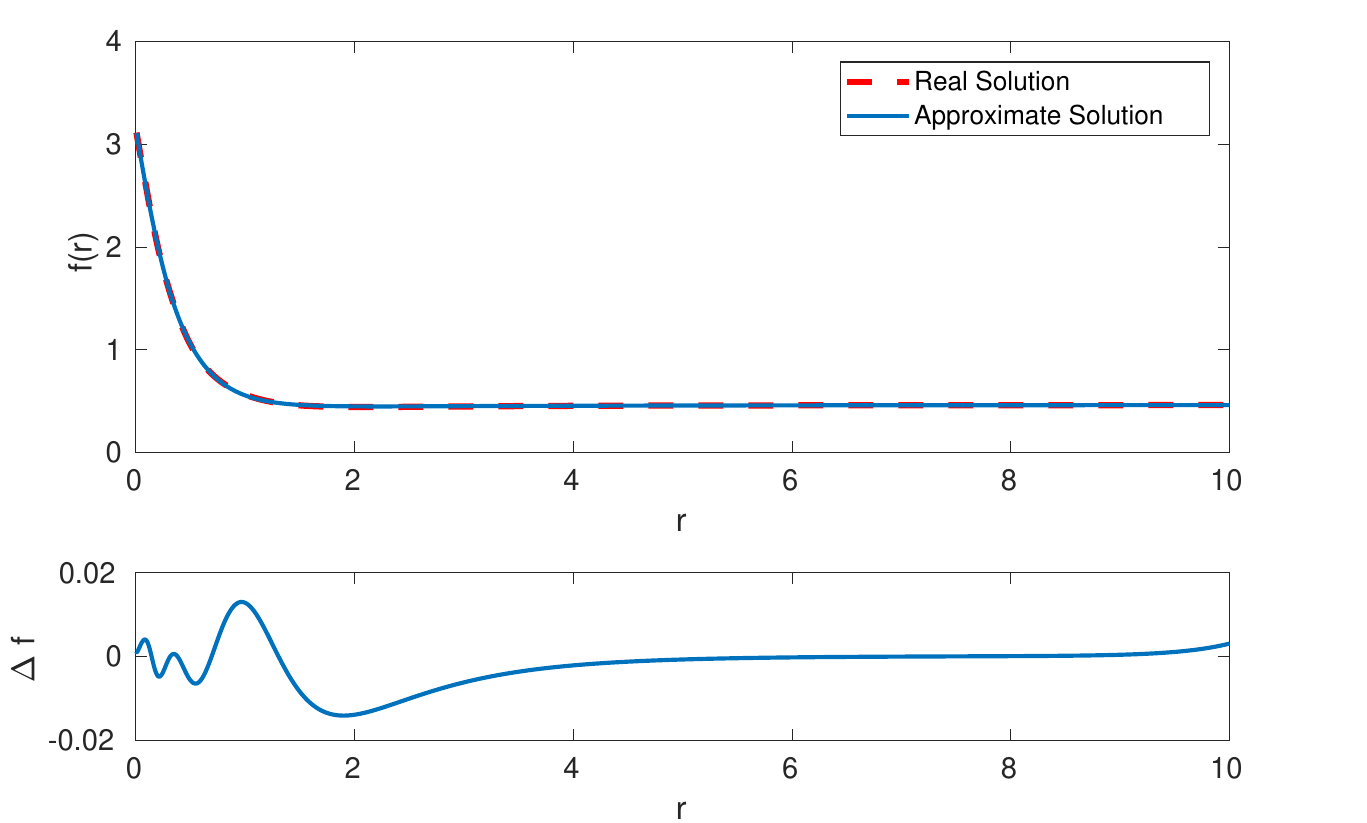}} 
\subfigure[]{\includegraphics[width=.48\linewidth]{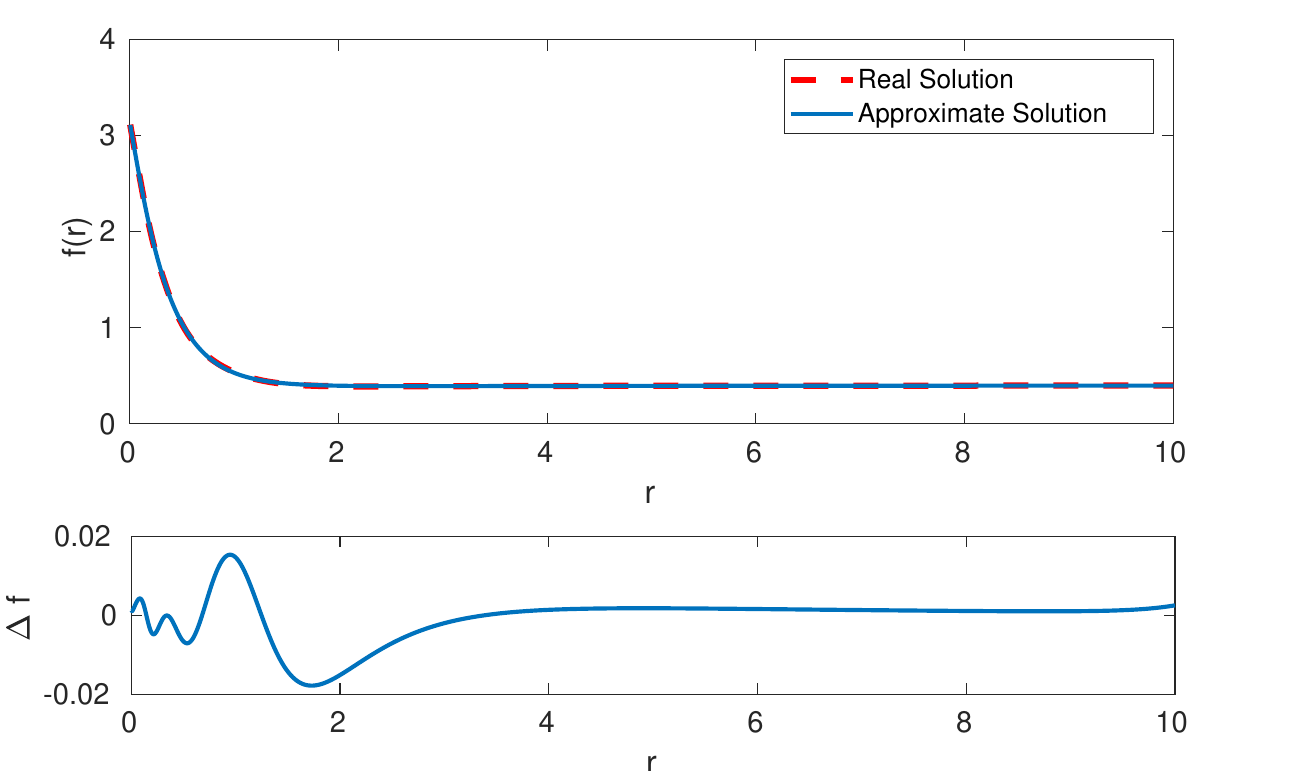}}
\subfigure[]{\includegraphics[width=.48\linewidth]{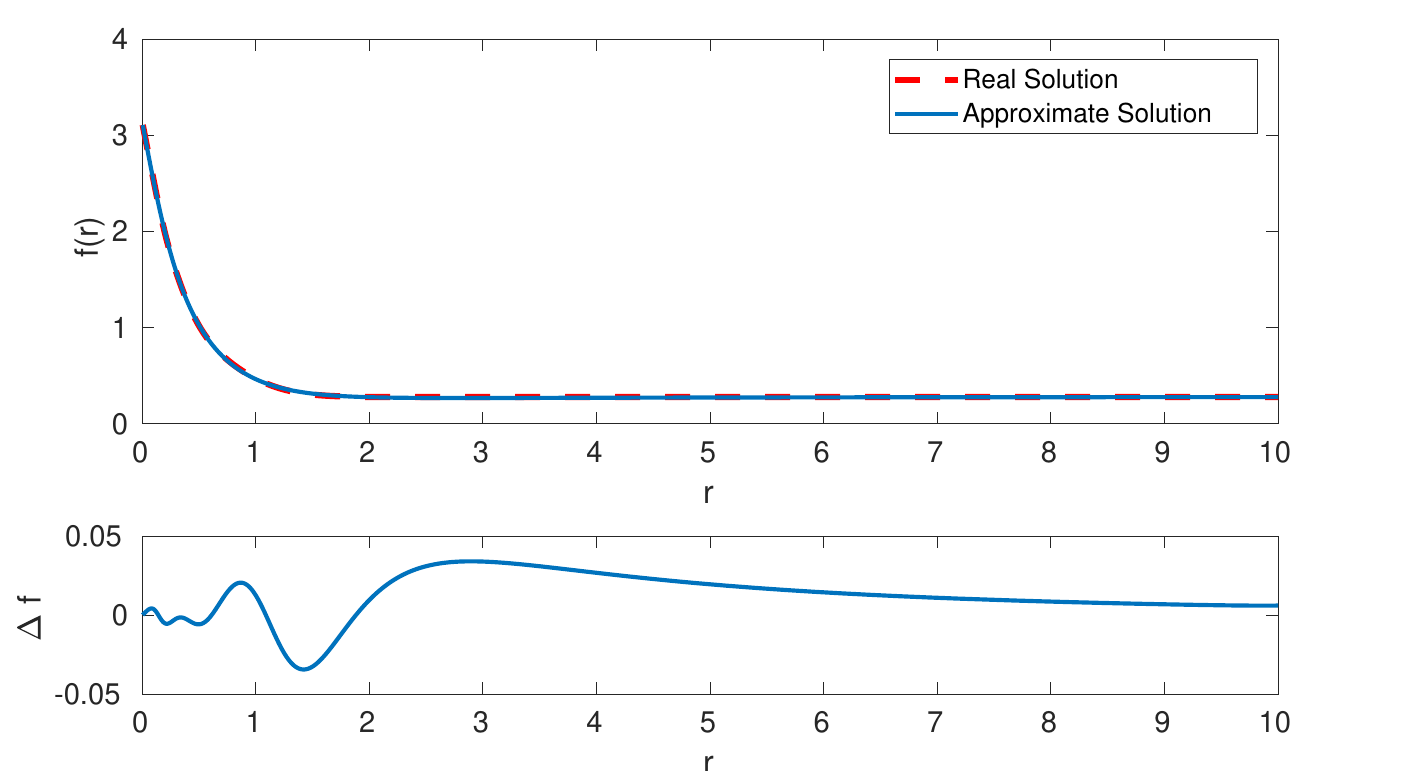}}
\subfigure[]{\includegraphics[width=.48\linewidth]{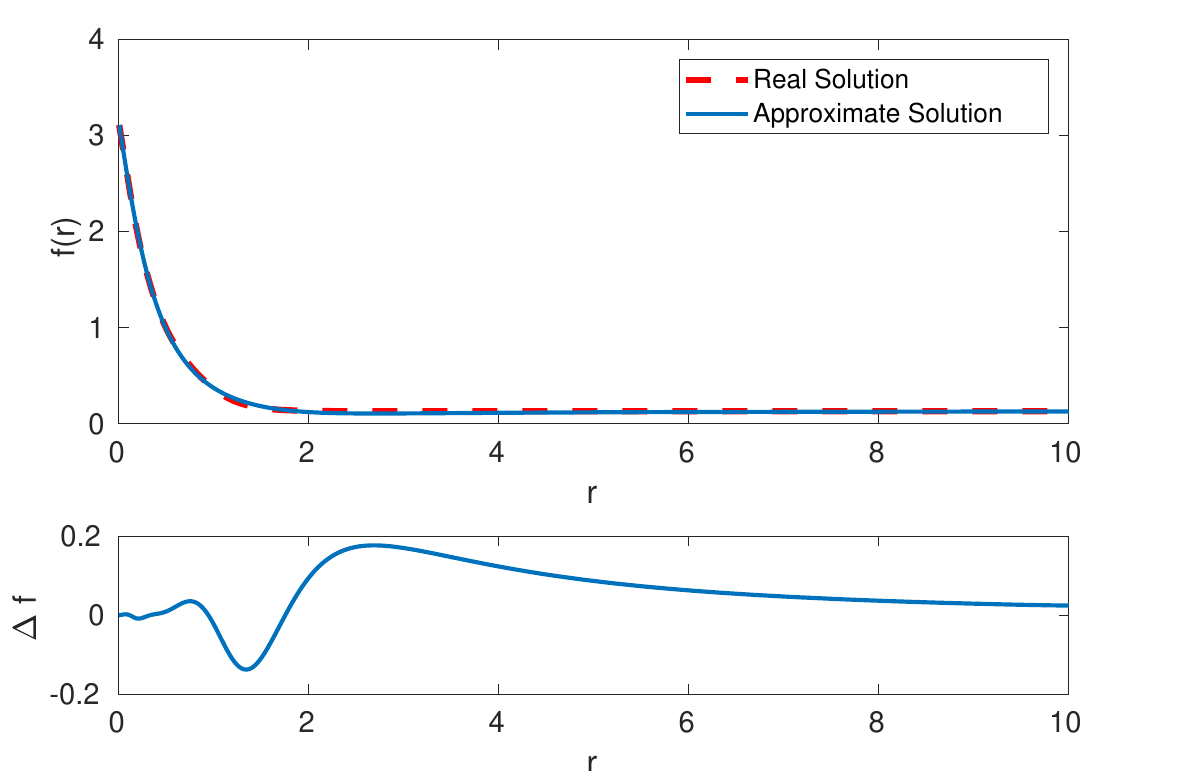}}
\subfigure[]{\includegraphics[width=.48\linewidth]{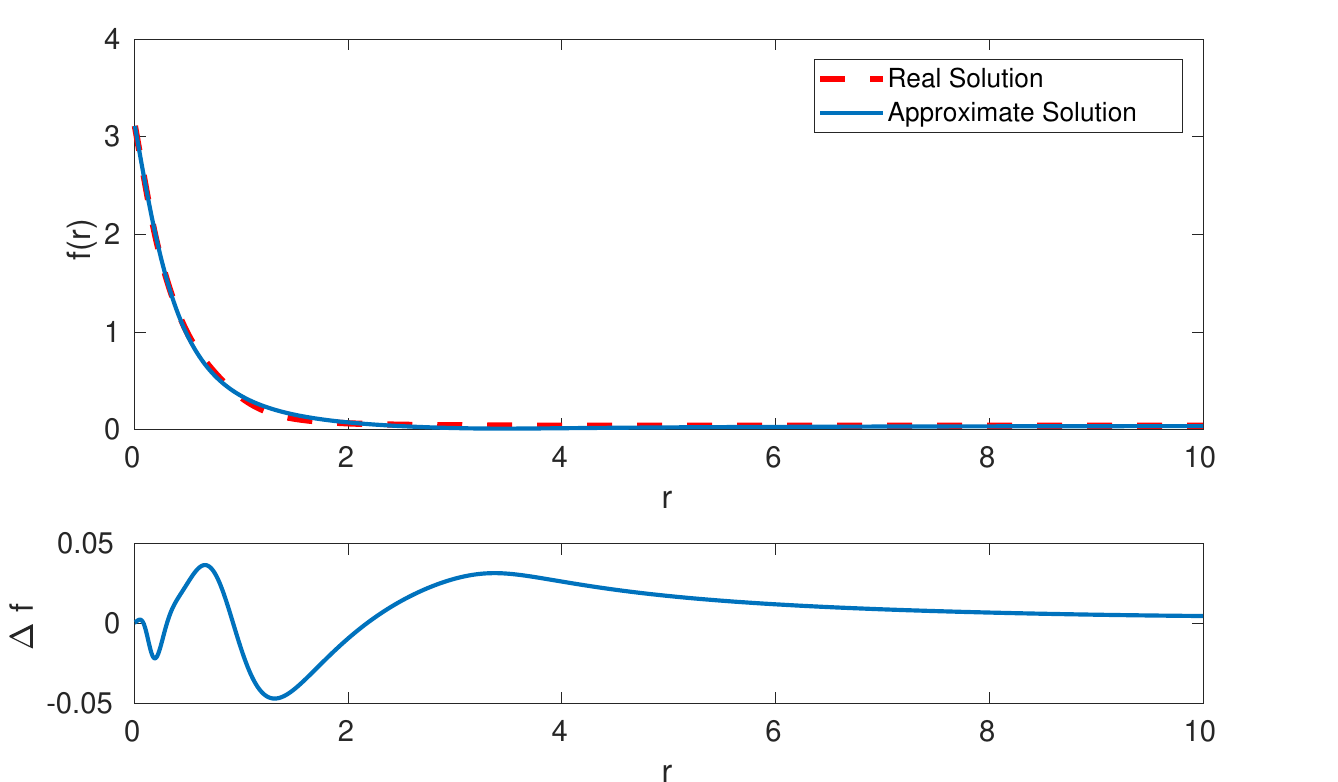}}
\caption{Best fitting results for $k_1= k_2=k_4=k_5=k_6=1, \; k_3=-3$ (the standard set) and $H/H_{\text{cr}}=\text{(a) }0,\;\text{(b) }0.3,\;\text{(c) }0.6,\;\text{(d) }0.9,\;\text{(e) }1$ for the case $f(0)=\pi$. The red dashed line represents the numerical solution to equation \eqref{skyeq} and the blue one represents the best fitting curve \eqref{fpifit}. On the bottom of each figure, the function $f(r)-f_a(r)$ is reported as an estimation of the goodness of the fit.}
\label{fig:fitpi}
\end{figure}

\begin{table}
\centering
\begin{tabular}{c|c|c|c}
$H/H_{\text{cr}}$ & $\overline{a} $ & $\overline{c}$ & $\overline{d}$ \\\hline
0				  & 0.7987  & 0.7639 & 0.0789 \\
0.3				  &  0.7999 &  0.7782  & 0.0794 \\
0.6				  &  0.9674  &0.9416 & 0.0959 \\
0.9               & 0.7903  &  0.7677  & 0.0788 \\
1				  & 	  0.4297  &  0.4270 &   0.0448\\
\end{tabular}
\caption{Best fitting parameters for $k_1= k_2=k_4=k_5=k_6=1, \; k_3=-3$ and $H/H_{\text{cr}}=0,\;0.3,\;0.6,\;0.9,\; 1$ for the case $f(0)=\pi$.}
\label{tab:fitpi}
\end{table}
In conclusion, our analysis leads to providing a good (within a few percentage) global approximated analytical expression of the Skyrmion tubes by using only 5 constants:
$\theta_0$, $\overline{a}, \overline{c}, \overline{d}$ and $\beta$. The physical meaning of $\theta_0$ and $\beta$ is straightforward and they can be measured by suitable experiments.
On the other hand, the remaining constants provide information about the shape of the Skyrmion tube which can be also experimentally observed.


\end{document}